\documentclass[a4paper,fleqn,usenatbib]{mnras}
\usepackage{newtxtext,newtxmath}

\usepackage[T1]{fontenc}
\usepackage{ae,aecompl}
\usepackage{subfigure}

\usepackage{graphicx}	
\usepackage{amsmath}	
\usepackage{amssymb}	
\usepackage{pdflscape}

\newcommand{\mg}{MG~0414+0534}
\newcommand{\rxj}{RXJ~0911+0551}
\newcommand{\qsob}{B~1422+231}
\newcommand{\wfi}{WFI~J2026-4536}

\title[HST imaging of four lensed quasars]{HST imaging of four gravitationally lensed quasars}

\author[N.F. Bate et al.]
{\parbox{\textwidth}{N. F. Bate,$^{1}$\thanks{E-mail: nbate@ast.cam.ac.uk (NFB)}
G. Vernardos,$^{2}$
M. J. O'Dowd,$^{3,4,5}$
D. M. Neri-Larios,$^{6}$
R. L. Webster,$^{6}$
D. J. E. Floyd,$^{6,7,8,9}$
R. L. Barone-Nugent,$^{6}$
K. Labrie,$^{10}$
A. L. King,$^{6}$
and S.-Y. Yong$^{6}$}\vspace{0.4cm}\\
\parbox{\textwidth}{$^{1}$Institute of Astronomy, University of Cambridge, Madingley Road, Cambridge CB3 0HA, UK\\
$^{2}$Kapteyn Astronomical Institute, University of Groningen, PO Box 800, NL-9700AV Groningen, the Netherlands\\
$^{3}$Department of Physics and Astronomy, Lehman College, City University of New York, 250 Bedford Park Boulevard West, Bronx, NY 10468-1589, USA\\
$^{4}$Department of Astrophysics, American Museum of Natural History, Central Park West and 79th Street, NY 10024-5192, USA\\
$^{2}$The Graduate Center of the City University of New York, 365 Fifth Avenue, New York, NY 10016, USA\\
$^{6}$School of Physics, The University of Melbourne, Parkville, VIC 3010, Australia\\
$^{7}$School of Physics, Monash University, Clayton, VIC 3800, Australia\\
$^{8}$Monash Centre for Astrophysics, Monash University, Clayton, VIC 3800, Australia\\
$^{9}$Think Big Analytics\\
$^{10}$Gemini Observatory, Hilo, HI 96720, USA}
}

\date{Accepted 2018 June 28. Received 2018 June 28; in original form 2017 May 18}

\pubyear{2017}

\begin{document}
\label{firstpage}
\pagerange{\pageref{firstpage}--\pageref{lastpage}}
\maketitle

\begin{abstract}
We present new HST WFPC3 imaging of four gravitationally lensed quasars: \mg;~\rxj;~\qsob;~\wfi. In three of these systems we detect wavelength-dependent microlensing, which we use to place constraints on the sizes and temperature profiles of the accretion discs in each quasar. Accretion disc radius is assumed to vary with wavelength according to the power-law relationship $r\propto \lambda^p$, equivalent to a radial temperature profile of $T\propto r^{-1/p}$. The goal of this work is to search for deviations from standard thin disc theory, which predicts that radius goes as wavelength to the power $p=4/3$. We find a wide range of power-law indices, from $p=1.4^{+0.5}_{-0.4}$ in \qsob~to $p=2.3^{+0.5}_{-0.4}$ in \wfi. The measured value of $p$ appears to correlate with the strength of the wavelength-dependent microlensing. We explore this issue with mock simulations using a fixed accretion disc with $p=1.5$, and find that cases where wavelength-dependent microlensing is small tend to under-estimate the value of $p$. This casts doubt on previous ensemble single-epoch measurements which have favoured low values using samples of lensed quasars that display only moderate chromatic effects. Using only our systems with strong chromatic microlensing we prefer $p>4/3$, corresponding to shallower temperature profiles than expected from standard thin disc theory.
\end{abstract}




\section{Introduction}
Observations of gravitationally microlensed quasars offer a unique opportunity to study the structure of quasar accretion discs at rest frame ultraviolet (UV) and optical wavelengths. These analyses hinge on the fact that gravitational lensing is achromatic, but the magnitude of any microlensing-induced brightness fluctuations depends strongly on the projected size of the emission region in the lensed source. In a quasar accretion disc, hotter regions are expected to be physically smaller than cooler regions, and so will be more strongly magnified by microlensing. 

Microlensing analyses have already thrown up challenges to the standard thin disc theory. It has been robustly established that quasar accretion discs in the observed optical ($2500$\AA~rest frame) are larger by factors of 2 to 4 than expected. This result is obtained from both single-epoch observations, and light curve analyses. In the latter case (e.g. \citealt{morgan+10}; the review of \citealt{chartas+16} and references therein), observationally-expensive monitoring campaigns are conducted to gather microlensing light curves, typically focused on accretion disc sizes at single, or occasionally two, wavelengths (usually X-ray and optical). Single-epoch studies (e.g. \citealt{pooley+07}; \citealt{blackburne+11}; \citealt{jv+12}) allow for easier analysis of larger samples of objects, and return similar results.

When multi-wavelength data is available, quasar microlensing observations can be used to map the radial profile of the accretion disc as a function of wavelength (\citealt{anguita+08}; \citealt{bate+08}; \citealt{eigenbrod+08}; \citealt{floyd+09}; \citealt{blackburne+11}; \citealt{blackburne+14}; \citealt{jv+14}; \citealt{rojas+14}; \citealt{macleod+15}). Typically, a power-law of the form $r\propto\lambda^p$ is assumed. The standard \citet{ss73} disc sets $p=4/3$ (hereafter SS), but other models predict different values for $p$ (e.g. \citealt{abramowicz+88}, $p=2.0$; \citealt{agol+00}, $p=8/7$). 

Under the assumption that the emergent accretion disc spectrum is the superposition of blackbody spectra generated locally in the disc, the spectral profile $r\propto\lambda^p$ is related to the temperature profile via the power law index $p$: $T\propto r^{-1/p}$ (see e.g. \citealt{frank+02}). Constraints on the power law index of the spectral profile are therefore also constraints on the radial temperature profile. If $p>4/3$, temperature falls off as a function of radius more slowly than expected from the SS disc (a shallower temperature profile). If $p<4/3$, the temperature profile is steeper, cooling down more rapidly with radius than the SS disc.

Early temperature profile measurements using microlensing have painted a conflicting picture. Single-epoch observations of microlensed accretion discs seem to favour steeper temperature profiles than SS ($p<4/3$, e.g. \citealt{floyd+09}; \citealt{blackburne+11}; \citealt{munoz+11}; \citealt{jv+14}), although there are exceptions (e.g. \citealt{bate+08}; \citealt{rojas+14}). This is difficult to reconcile with measurements of larger than expected discs at optical wavelengths, since a steeper temperature profile implies that the accretion disc will be smaller at a given wavelength, not larger.

Most temperature profile measurements have been performed on single systems, or occasionally a handful. The most recent ensemble measurement is \citet{jv+14}, hereafter JV14, who analysed ten image pairs in eight lensed quasars. They found a joint Bayesian estimate for the power law index of $p=0.8\pm0.2$ at 68 per cent confidence, well below the fiducial $p = 4/3$. 

Gravitational microlensing is currently the only technique that has been used to study the physical structure of these objects at high redshifts ($z\sim1.5$ to $3.5$). Photometric reverberation mapping has been used to probe accretion discs in lower-redshift active galactic nuclei. The AGN Space Telescope and Optical Reverberation Mapping project has examined the accretion disc in the Seyfert NGC 5548 in considerable detail (see especially \citealt{fausnaugh+16}; \citealt{starkey+17}), using 19 overlapping continuum light curves to measure a larger size and steeper temperature profile ($p=1.01\pm0.03$) in that system. A similarly-detailed analysis of NGC 4593 \citep{cackett+17} also found a larger than expected size, but the measured temperature profile was consistent with the standard thin disc prescription. \citet{jiang+17} analysed a sample of 240 $z\sim0.1$ to $0.3$ quasars from Pan-STARRS. They again measure larger accretion disc sizes than expected from thin-disc theory, however they favour flatter temperature profiles (with the caveat that their wavelength coverage is much shorter than the \citealt{starkey+17} or \citealt{cackett+17} analyses).

Understanding the diversity of temperature profile measurements, along with the apparent contradiction between larger discs and (possibly) steeper temperature profiles requires more data, and careful consideration of sources of error and contamination in both our observations and our analysis techniques. The principle contaminants in single-epoch observations are: time delays between lensed images causing us to catch the background quasar in different underlying states; broad emission lines that fall in broadband filters and dilute the signal from the quasar continuum; and differential extinction. Possible systematic effects have been poorly explored.

In this paper, we present multi-wavelength Hubble Space Telescope (HST) observations of four gravitationally lensed quasars: \mg~\citep{hewitt+92}, \rxj~\citep{bade+97}, \qsob~\citep{patnaik+92}, and \wfi~\citep{morgan+04}. These observations were tuned to mitigate some of the major challenges and sources of uncertainty facing multi-band single-epoch imaging analyses.

In Section \ref{sec:lenses} we discuss issues around our sample selection, including common contaminants in single-epoch microlensing analyses. We also present background detail on the four lensed quasars studied here, summarising any existing microlensing analyses. In Section \ref{sec:data} we present our data, and describe the reduction process. In Section \ref{sec:sims} we lay out our simulation technique. First we discuss macro-lens models for each of our systems, then the single-epoch microlensing analysis technique. This technique broadly follows JV14, but differs in a few key details. In Section \ref{sec:results} we present the results of our analysis, and then discuss their implications in Section \ref{sec:discuss}. This discussion leads to a suite of mock observations in Section \ref{sec:mocks}, which clarify the origin of the steep temperature profiles measured in JV14. Finally, we conclude in Section \ref{sec:conclusions}.

Throughout this paper we use a cosmology with $H_0 = 72\ \rmn{kms}^{-1}\ \rmn{Mpc}^{-1}$, $\Omega_\rmn{m} = 0.3$ and $\Omega_\Lambda = 0.7$.

\section{Sample Selection}
\label{sec:lenses}

In this section, we discuss common issues relating to the single-epoch imaging technique, including methods for their mitigation. We then present our HST sample of four gravitationally lensed quasars, and briefly discuss any previous microlensing analyses of these systems.

\subsection{Single-Epoch Imaging Technique Contaminants}

Single-epoch imaging studies of quasar microlensing are subject to a number of potential contaminants, which have been successfully handled to a greater or lesser degree. The principle contaminants in the observations are: (i) time delays between lensed images; (ii) broad emission lines that fall in broadband filters and dilute the signal from the quasar continuum; and (iii) differential extinction. Finally, (iv) possible systematic effects in our analysis technique have been poorly explored. We will discuss each of these issues in turn, focussing on how we have mitigated them in the current analysis.

\begin{enumerate}
\item Quasars are variable on all timescales longer than a day \citep{macleod+12}, and so time delays between lensed images can mean that the background quasar is captured in a different state in each image with a single observation. In order to eliminate quasar variability as a source of contamination, we work only with close image pairs. In these cases, time delays are expected to be short (less than a day), and so we can be confident that the quasar is in the same state in each image.

Close image pairs provide an ideal laboratory for microlensing constraints on quasar accretion discs (\citealt{bate+07}; \citealt{bate+08}; \citealt{floyd+09}). In the presence of a high smooth matter fraction -- expected in the outskirts of lensing galaxies, where lensed quasar images typically form -- one of the close images (the saddle point image) is preferentially suppressed by microlensing (\citealt{schechter+02}; \citealt{vernardos+14}). This frequently leads to a large magnification difference between the close images, which produce the tightest single-epoch accretion disc size constraints. It is often overlooked that, conversely, small magnification differences between lensed images provide very little information on accretion disc sizes (see Figure 4 in \citealt{bate+07}).

With their $<0.5\arcsec$ separation, deblending of close image pairs is a significant challenge with ground-based imaging. For this reason, the photometric precision afforded by HST is essential.

\item Measurements of the continuum slope with broad multi-band photometry risk line contamination, which arises due to differing physical scales for broad line and continuum emission in the source quasar. Since broad lines are expected to be emitted from larger physical scales than the continuum, they will be affected by microlensing differently (see e.g. \citealt{abajas+02}; \citealt{wayth+05}; \citealt{odowd+11}; \citealt{sluse+12b}). If a broadband filter overlaps a broad emission line, the flux you observe in that filter is therefore a combination of broad line and continuum emission, and so convolves microlensing information on two physical scales. 

Line contamination can be avoided with sufficient spectral resolution. However, this does not necessitate full spectroscopy. Narrow- or medium-band filters can be chosen to cleanly select regions of the quasar spectrum free from broad line contamination (e.g. \citealt{mosquera+11a}). This approach grants higher signal-to-noise for a given exposure time than spectroscopy, and also greatly simplifies the deblending of close lensed image pairs. This is the approach we have chosen, using medium-band HST filters to minimise the impact of broad emission lines on our results.

\item Differential extinction can produce chromatic variation that mimics the effects of microlensing (see e.g. \citealt{odowd+18}). This can be effectively removed by monitoring the system and conducting light curve analyses: differential extinction is not expected to vary on microlensing-like timescales. 

Accounting for it in single-epoch observations is much harder. In the ideal case, we would have observations of regions in the quasar that are sufficiently large to be unaffected by microlensing, emitting at wavelengths very similar to the continuum, to establish the baseline impact of differential extinction. For example, \citet{mediavilla+09} and JV14 use the narrow component of broad emission lines to estimate intrinsic flux ratios. In the absence of adequate spectroscopy, infrared or radio flux ratios can be used to estimate intrinsic magnification ratios. This is the method we adopt here.

\item The final issue is systematics in our simulation technique, and it remains relatively unexplored. Single-epoch analyses have proceeded under the assumption that the simulations provide a clean measurement of the structure of the quasar accretion disc, if contaminants in the observational data are minimised. In this paper, we will present the first evidence that this assumption may not be strictly true.

This last issue will be particularly important in the forthcoming synoptic survey era. Currently, we have temperature profile measurements for only a handful of lensed quasars; only $\sim100$ are known (see \citealt{mosquera+11} for a compilation). The Large Synoptic Survey Telescope (LSST,  \citealt{lsst+09}) alone is expected to discover thousands more \citep{oguri+10}, enabling studies of true statistical samples. Future surveys may provide us with light curves of these systems, superseding the single-epoch technique. However, the timescales for microlensing variations are often years or decades, and so there will still be a use for single-epoch analyses for the foreseeable future.
\end{enumerate}

\subsection{HST Sample}

The four lensed systems studied here were specifically chosen from a larger program to have no obvious ring features in our data. The presence of full or partial Einstein rings indicates that the quasar host galaxy is being significantly lensed. Rings or arcs provide more constraints for modelling the lens, however they may be an additional source of contamination in any measured flux ratios between lensed images. This issue will be explored in more detail in subsequent papers.

\subsubsection*{\mg}

This system, first reported in \citet{hewitt+92}, has been the subject of a significant number of lensing analyses. This is in part due to an anomaly in the flux ratio between images $A_2$ and $A_1$ that persists into the infrared and the radio. This has been taken to indicate the presence of unobserved substructure in the lens, on so-called millilensing scales (approximately dwarf galaxy-sized, see e.g. \citealt{mao+98}; \citealt{dalal+02}; \citealt{kochanek+04}; \citealt{macleod+13}).

In \citet{pooley+07}, the authors used X-ray and optical data to study the relative sizes of the accretion discs in 10 systems, including \mg. They found that the optical continuum emission in this quasar arose from a region $\sim3$ times larger than expected from thin disc theory. \citet{blackburne+11} followed up this analysis by also constraining the temperature profile of the accretion disc, finding a value consistent with the SS $p=4/3$ scaling. It is worth noting that this result was at odds with all of the other systems in their paper, for which they found very low values of $p\sim0.2$. An X-ray microlensing size measurement of \mg~also appears in \citet{jv+15b}.

We have also previously studied the accretion disc in \mg~(\citealt{bate+08}; \citealt{bate+11}). Although the errors in our previous analysis were large, we found an accretion disc temperature profile formally consistent with the SS disc at the $2\sigma$ level, with a preference for power law indices $p>4/3$ (formal constraint: $1.6^{+0.6}_{-0.5}$ at 68 per cent confidence). Our analysis was conducted with an earlier version of the single-epoch imaging technique, which didn't account for errors in macro-modelling. Given the possible presence of substructure near the $A_2/A_1$ image pair, this omission may be significant.

\subsubsection*{\rxj}

\rxj~is an X-ray bright lensed quasar originally detected in the ROSAT All-Sky Survey \citep{bade+97}. It has appeared in the X-ray accretion disc analyses of \citet{pooley+07}, \citet{blackburne+11}, and \citet{jv+15b}. In \citet{pooley+07}, the quasar was found to have an optical accretion disc size $\sim6$ times larger than expected from thin disc theory. 

Using optical, IR and X-ray data, \citet{blackburne+11} reported a temperature profile for this system consistent with no wavelength dependence ($p = 0.17\pm0.41$). These particularly low values of $p$ (corresponding to steep temperature profiles) were the overall result for their analysis of 12 systems.

\subsubsection*{\qsob}

\qsob~\citep{patnaik+92} is also part of the \citet{pooley+07} sample of lensed quasars. It differs from \mg~in that its optical size measurement is apparently consistent with thin disc theory. It is also included in the joint size analyses of \citet{mediavilla+09}, \citet{jv+12}, \citet{jv+15a} and \citet{jv+15b} (the first three using optical data, the latter X-ray), but to date no temperature profile measurement has been reported. 

\citet{mao+98} used anomalies in the radio flux ratios to argue for the presence of substructure in the lens. This was the first time that this use for lensed quasar observations was explicitly discussed in the literature. Our lens models are discussed in Section \ref{subsec:models}; for this system, we make no attempt to account for millilensing substructure.

\subsubsection*{\wfi}

\wfi~is the most recently discovered system in this paper \citep{morgan+04}, and has only appeared in one microlensing accretion disc analysis: \citet{blackburne+11}. They again found an extremely steep temperature profile, consistent with no wavelength dependence ($p = 0.27\pm0.53$).

Additionally, \citet{sluse+12} reported that \wfi~is a problematic system to model. This will be discussed further in Section \ref{subsec:models}. The system's lens redshift is currently unknown.

\section{Data}
\label{sec:data}

\begin{figure}
  \includegraphics[width=85mm]{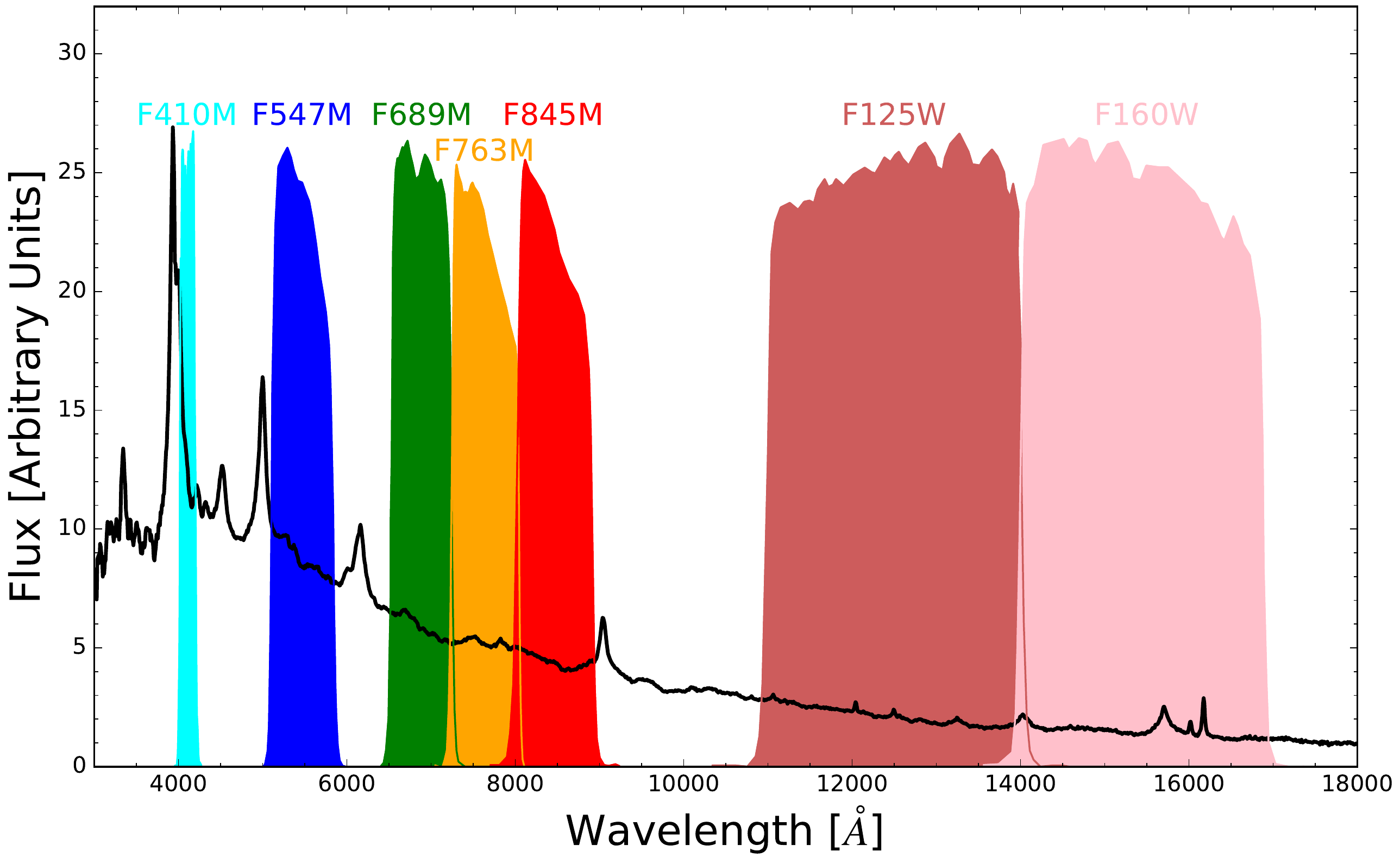}
  \caption{HST WFC3 filter transmission curves, overlaid on a quasar composite spectrum \citep{vandenberk+01} at the redshift of \wfi~($z=2.23$). This illustrates how we have used medium-band filters to select regions between broad emission lines, to avoid contaminating the signal from the microlensed accretion disc.}
  \label{fig:2026filters}
\end{figure}

Observations were taken in HST Cycle 20 (Program ID 12874, PI Floyd). Imaging was performed with HST's Wide Field Camera 3 (WFC3) in both the UVIS and IR channels, utilizing four to seven medium-band filters for each target. Medium-band imaging allow us to select filters that fall between broad emission lines. This is important because the broad emission line region is much larger than the accretion disc and so experiences different microlensing magnification (when it is microlensed at all). The suite of WFC3 medium band filters allows us to avoid these lines at all source redshifts. Figure~\ref{fig:2026filters} shows the example case of \wfi: filter transmission curves are overlaid on a composite quasar spectrum \citep{vandenberk+01} at this quasar's redshift. Table~\ref{obsdetail} details exposure times, dates and filters for each source. 

Data were reduced using the standard HST Astrodrizzle reduction pipeline. Fluxes of individual lensed images were measured using GALFIT \citep{peng+02} to fit a point spread function (PSF) at the location of each lensed image, while the lensing galaxy was fit with a Sersic profile. Free parameters in the fit included lensed image and lensing galaxy fluxes as well as positions of all sources. TinyTim \citep{krist+11} was used to generate WFC3 PSFs, and StarFit (Timothy Hamilton, private communication) was used to propagate these PSFs through the dither-combine pipeline to match the final combined PSFs. Table \ref{tab:fluxes} gives the measured magnitudes for all lensed images and lensing galaxies. 

\begin{table*}
\caption{HST WFC3 observation details}
\label{obsdetail}
\begin{tabular}{lcccccc}
\hline          
System & $z_S$ & $z_L$ & Obs. Date & Orbits & WFC3/UVIS & WFC3/IR\\
\hline          
\mg & 2.64 & 0.96 & 2013-07-25 & 2 & F763M F845M & F125W F160W\\
\rxj & 2.80 & 0.77 & 2012-10-19 & 2 & F547M F621M F689M F845M & F125W F160W\\
\qsob & 3.62 & 0.34 & 2013-07-12 & 2 & F621M F763M F845M & F105W F125W F160W\\
\wfi & 2.23 & ---  & 2012-11-22 & 2 & F410M F547M F689M F763M F845M & F125W F160W\\
\hline   
\end{tabular}
\end{table*}

In Table \ref{tab:ratios} we present observed magnitude differences for the image pairs of interest (typically the close image pair). The observed wavelengths are taken to be the pivot wavelengths of each filter. These are the data that are used for the microlensing analysis presented in this paper.

\section{Modelling and Simulations}
\label{sec:sims}

In this section, we describe the various components of our lens modelling and microlensing simulation technique.

\subsection{Lens models}
\label{subsec:models}

Models of the lensing potentials in each system provide us with the key parameters for microlensing simulations: the convergence $\kappa$ and shear $\gamma$ at each lensed image position. 

\subsubsection*{\mg}
\mg~is known to harbour substructure in the vicinity of the anomalous $A_1$ and $A_2$ image pair. Simple lens models consisting only of a singular isothermal ellipse and external shear do not reproduce the $A_2/A_1$ flux ratios observed in the radio and mid-IR (e.g \citealt{minezaki+09}; \citealt*{katz+97}). \citet{macleod+13} used these data to search for the best-fitting location for a dark substructure near the anomalous image pair. They found a best-fitting mass of $10^{6.2}M_{\sun}$ to $10^{7.5}M_{\sun}$, located to the north-east of image $A_2$ (modelled as a singular isothermal sphere at the redshift of the lensing galaxy).

We take the \citet{macleod+13} G1+G2+G3 model using both mid-IR and radio data (see their Table 3, repeated here in Table \ref{tab:lensmodel}), and calculate convergences and shears at the $A_1$ and $A_2$ image positions using \textsc{lensmodel}\footnote{\url{http://physics.rutgers.edu/~keeton/gravlens/2012WS/}} \citep{keeton01}. This model includes three lenses: the primary lens galaxy G1, a secondary nearby (observed) galaxy G2, and the (dark) substructure G3 discussed in the previous paragraph. The microlensing parameters for this model are provided in Table \ref{tab:macromodels}.

\subsubsection*{\rxj~and \qsob}
For \rxj~and \qsob~we use lens models from the literature \citep{schechter+14}. These models consist of a singular isothermal ellipsoid (SIE) with an orientation and ellipticity constrained to match the observed shape of the stellar component in each lensing galaxy, plus an external shear term (referred to as SIE+X or SIE+$\gamma$ models). The \rxj~model also includes a secondary lens at the position of an observed smudge in the system, modelled as a singular isothermal sphere \citep{blackburne+11}. The model parameters are reproduced in Table \ref{tab:lensmodel} for convenience, and their resulting convergences and shears are summarised in Table \ref{tab:macromodels}.

\begin{table*}
\centering
  \caption{Lens model parameters} \label{tab:lensmodel}
  \begin{tabular}{lcccccccccccl} 
  \hline
    & \multicolumn{3}{c}{Primary Lens} & \multicolumn{2}{c}{Shear} & \multicolumn{3}{c}{Secondary Lens} & \multicolumn{3}{c}{Tertiary Lens} & \\
 Object & $\theta_{Ein}$ & $e$ & $\phi_e$ & $\gamma$ & $\phi_{\gamma}$ & $b_2$ & $x_2$ & $y_2$ & $b_3$ & $x_3$ & $y_3$ & Source \\
 \hline
 \mg & $1\farcs084$ & $0.238$ & $-83.9$ & $0.094$ & $53.1$ & $0\farcs176$ & $0\farcs857$ & $0\farcs181$ & $0\farcs007$ & $-0\farcs97$ & $-1\farcs39$ &\citet{macleod+13} \\
 \rxj & $0\farcs95$ & $0.11$ & $-70.0$ & $0.294$ & $8.3$ & $0\farcs22$ & $-0\farcs754$ & $0\farcs665$ &--&--&--&\citet{schechter+14} \\ 
 \qsob & $0\farcs74$ & $0.39$ & $-58.9$ & $0.137$ & $-47.2$ &--&--&--&--&--&--&\citet{schechter+14} \\ 
 \wfi & $0\farcs651$ & $0.133$ & $-25.0$ & $0.133$ & $82.9$ &--&--&--&--&--&--&This work \\ 
 \hline
\end{tabular}
\end{table*}

\begin{table*}
\centering
  \caption{Microlensing parameters, convergence $\kappa$ and shear $\gamma$} \label{tab:macromodels}
  \begin{tabular}{lccccccl} 
  \hline
 System & Image & $\kappa$ & $\gamma$ & Image & $\kappa$ & $\gamma$ & Source \\
 \hline
 \mg & $A_1$ & 0.51 & 0.42 & $A_2$ & 0.56 & 0.51 & \citet{macleod+13} \\ 
 \rxj & $A$ & 0.65 & 0.54 & $B$ & 0.59 & 0.28 & \citet{schechter+14} \\
 \qsob & $A$ & 0.38 & 0.47 & $B$ & 0.49 & 0.63 & \citet{schechter+14} \\
 \wfi & $A_1$ & 0.52 & 0.40 & $A_2$ & 0.54 & 0.54 & This work \\
 \hline
\end{tabular}
\end{table*}

\subsubsection*{\wfi}
\wfi~is a slightly more complicated case. Both \citet{chantry+10} and \citet{sluse+12} found this system difficult to fit accurately with simple lens models. In \citet{chantry+10}, they chose to constrain the ellipticity and position angle of an SIE+X lens model by the light (similar to the models in \citealt{schechter+14}). Imposing these constraints resulted in a large formal $\chi^2$; models with a significant offset between the position angles of the mass and the light were preferred. In \citet{sluse+12}, they relaxed these constraints and also added a second mass component to describe a nearby galaxy. In this case they again prefered models where the mass and light are significantly misaligned, possibly hinting at the presence of dark substructures.

In light of these complications, we opted to generate a simple SIE+X model where we left the orientation and the ellipticity of the lens as free parameters. We kept the position of the lens fixed to the centre of the light profile, and used only the image positions as constraints. The best fit model, generated with the \textsc{lensmodel} code, is provided in Table \ref{tab:lensmodel}. It is similar to the SIE+X case in \citet{sluse+12}. We emphasise that this model should be considered illustrative; it does not include the full complexities of the system.

The \wfi~lens redshift is also currently unknown. To convert our measured sizes into physical parameters we adopt a lens redshift of $z_{l,2026}=1.04$, estimated from the image separations and probability distributions for $z_l$ (\citealt{mosquera+11}; \citealt{ofek+03}).

\subsection{Simulation technique}
\label{subsec:technique}

We broadly follow the simulation technique laid out in JV14, with two important differences. The technique is described below, highlighting the areas where our methods differ. The basic outline is:

\begin{description}
\item \ref{subsubsec:maps}: Generate magnification maps from lens models.
\item \ref{subsubsec:isolate}: Isolate the microlensing signal from macrolensing using infrared or radio observations.
\item \ref{subsubsec:sims}: Conduct Bayesian microlensing simulations to extract accretion disc parameters.
\end{description}

\subsubsection{Magnification maps}
\label{subsubsec:maps}

The lens models described in the previous section provide the two key microlensing parameters at the location of each lensed image: the convergence $\kappa$ and shear $\gamma$ (see Table \ref{tab:macromodels}). The convergence can be split into two components: a clumpy component $\kappa_*$ that describes point-mass (stellar) microlenses, and a smooth component $\kappa_s$, such that $\kappa = \kappa_* + \kappa_s$. The smooth matter fraction is $s = \kappa_s / \kappa$. 

In the systems studied here, the lensed images lie in the outskirts of the lens galaxies, so we expect the smooth matter fraction to be high. Typical values obtained from previous microlensing analyses range from smooth matter fractions of $\sim80$ per cent (e.g \citealt{bate+11}; \citealt{jv+15a}) to $\sim 93$ per cent (e.g. \citealt{pooley+12}). JV14 took the smooth matter fraction to be 95 per cent in their analysis; we chose instead to vary it. We used 11 smooth matter fractions $s = 0.0, 0.1, 0.2, \ldots, 0.9, 0.99$, and discuss the impact of smooth matter fraction on our results below. This is the first key difference between our analysis and that of JV14.

Magnification maps were generated using the GPU-D code (\citealt{thompson+10}; \citealt{bate+12}), within the GERLUMPH framework (\citealt{vernardos+14}; \citealt{vernardos+15}). We used maps with a side length of 100 Einstein radii, and 10000 pixels, and $M_\rmn{micro} = 1M_{\sun}$ microlenses. This corresponds to physical per-pixel resolutions of 0.1445 (\mg), 0.1618 (\rxj), 0.2207 (\qsob), and 0.1341 (\wfi) light days. This resolution is $\sim0.5$ times the innermost stable circular orbit of a $10^9M_{\sun}$ black hole, sufficient for our rest-frame UV to optical observations. 

\begin{table*}
\centering
  \caption{Unmicrolensed baselines $\Delta m_\rmn{macro}$} \label{tab:baselines}
  \begin{tabular}{lccccl} 
  \hline
 System & Image Pair & Observed & Lens model & Adopted & Observed wavelength \\
 \hline
 \mg & $A_2-A_1$ & $0.09\pm0.02$ & $0.04$ & $0.09\pm0.05$ & 11.7 $\mu$m \citep{minezaki+09}; \\
 & & & & & 11.2 $\mu$m \citep{macleod+13} \\ 
 \rxj & $B-A$ & $-0.74\pm0.10$ & $-0.69$ & $-0.74\pm0.10$ &  5 GHz \citep{jackson+15} \\
 \qsob & $B-A$ & $-0.08\pm0.02$ & $-0.20$ & $-0.08\pm0.12$ &  8.4 GHz \citep{patnaik+99} \\
 \wfi & $A_2-A_1$ & -- & $0.14$ & $0.14\pm0.25$ & -- \\
 \hline
\end{tabular}
\end{table*}

All physical sizes quoted in this paper depend on the choice of microlens mass $M_\rmn{micro}$, so all physical units can be rescaled by a factor of $(M_\rmn{micro}/M_{\sun})^{1/2}$. We omit this factor from scales throughout this paper for readability, but it should be assumed to be present in every physical measurement obtained via microlensing simulations.

All magnification maps are available for download via the GERLUMPH website\footnote{\url{http://gerlumph.swin.edu.au}}.

\subsubsection{Isolating the microlensing signal}
\label{subsubsec:isolate}

We conduct our microlensing simulations using the magnitude differences $\Delta m_\rmn{obs} = m_B - m_A$ (or $m_{A_2} - m_{A_1}$ as appropriate) provided in Table \ref{tab:ratios}. We note that the simulations can trivially be run in terms of flux ratio instead, as in our previous papers (\citealt{bate+08}; \citealt{floyd+09}), using the relationship $\Delta m_\rmn{obs} = m_B - m_A = -2.5\rmn{log}_{10}(f_B/f_A)$.

In an ideal case, we would isolate the microlensing signal from both macrolensing and differential extinction by using a series of $\Delta m$ measurements from emission regions in the source large enough to be unaffected by microlensing. These emission regions would need to emit at similar wavelengths to the continuum source in order to accurately map the effects of extinction.

The procedure adopted by \citet{mediavilla+09} and following papers was to use the centres of broad emission lines in spectra to establish unmicrolensed baselines. We do not currently have access to suitable spectra for all of our systems, and in any case there is some question over the accuracy of the assumption that no microlensing is present in the centres of broad emission lines.

We chose instead to normalise all of our observed magnitude differences within a given system to a single unmicrolensed baseline: $\Delta m_\rmn{micro} = \Delta m_\rmn{obs} - \Delta m_\rmn{macro}$. The baseline $\Delta m_\rmn{macro}$ is taken in the radio or infrared where possible, and removes only the effects of macrolensing. This is the second key difference between our analysis and JV14. In practice, it means that we are attributing all of the chromatic variation in our observations to microlensing, and none to any other chromatic effects such as differential extinction. We note that since extinction can effect either image in a lensed pair, neglecting differential extinction could lead us to over-estimate \textit{or} under-estimate the size of any chromatic microlensing.

The literature data used to determine unmicrolensed baselines are provided in Table \ref{tab:baselines}. We can compare the observed baselines with the predictions from our lens models, to check for consistency. In two cases, \mg~and \qsob, we find that our lens model values lie marginally outside the observed values. To account for this, we expand the errors on the unmicrolensed baselines used in our simulations. We do not have suitable observations for \wfi, so we choose to assume that the lens model prediction for the unmicrolensed baseline is correct, but assign it a correspondingly large error (23 percent in the ratio). This value for the error was chosen conservatively: it is approximately twice the percentage error in the least-accurately measured unmicrolensed baseline in our sample (\rxj), and approximately twice the largest deviation between model predicted and observed values (\qsob).

In Table \ref{tab:dm_micro}, we provide the measured microlensing amplitudes $\Delta m_\rmn{micro} = \Delta m_\rmn{obs} - \Delta m_\rmn{macro}$ for each of our four systems. Here we have simply added the errors in the unmicrolensed baseline (Table \ref{tab:baselines}, fifth column)  to the observational errors in quadrature. We note that these errors are handled differently in our microlensing simulations, as they represent a systematic offset rather than an independent error in each data point. This will be described in detail in the next section. Table \ref{tab:dm_micro} is provided to give a sense of the size of the microlensing signal in each system.

\subsubsection{Microlensing simulations}
\label{subsubsec:sims}
We use a Bayesian analysis to constrain the size of the quasar accretion disc $r_s$ at $\lambda_0 = 1026$\AA, and the power law index $p$. The radial profile of the accretion disc is assumed to be of the form:

\begin{equation}
\label{eqn:profile}
r = r_s \left(\frac{\lambda}{\lambda_0}\right)^p.
\end{equation}
We use Gaussian profiles to describe the shape of the disc at each wavelength; the measured sizes $r_s$ are Gaussian dispersions. \citet*{mortonson+05} have demonstrated that size estimates from microlensing simulations are independent of the detailed surface brightness profile shapes, but rather depend only on the characteristic width of that profile.

Following JV14, we vary parameters on a regular grid, such that $ln(r_s) = 0.3 \times j$ for $j = 0 \ldots 11$, and $p = 0.25\times i$ for $i = 0 \ldots 15$. This is a slightly larger range in $p$ than in JV14. 

For each combination of $ln(r_s)$, $p$, and $s$ the magnification maps for image 1 and image 2 are convolved with a series of Gaussian source profiles with dispersion $r$ calculated at the rest-frame wavelengths of the observations using Equation \ref{eqn:profile}. Each map is then sampled on a regular grid of $10^4$ points. This allows us to construct $10^8$ simulated magnification ratios per $ln(r_s)$-$p$-$s$ parameter combination for comparison with the observational data. We note that we sample only from the central $2000\times 2000$ pixel area of each magnification map; $10000 \times 10000$ pixel maps allow us to convolve with large source profiles while avoiding convolution edge effects. The effective upper limit on our simulated source sizes is 8 Einstein radii.

Errors in the unmicrolensed baseline affect observed flux ratios at each wavelength equivalently: they simply act as a systematic offset. To account for this, we include the unmicrolensed baseline $\Delta m_\rmn{macro}$ as a nuisance parameter in our Bayesian analysis. At each of the $10^8$ simulated observations, we re-sample $\Delta m_\rmn{macro}$ from a Gaussian with mean and dispersion equal to the observed unmicrolensed baseline and its error (see Table \ref{tab:baselines}), and re-normalise the observed data. This effectively marginalises over errors in the unmicrolensed baseline.

We note that this differs from the usual technique (e.g. JV14), where the error in the unmicrolensed baseline is simply added into the observed errors in quadrature. This is equivalent to assuming that errors in the baseline affect each wavelength independently, rather than systematically.

We construct likelihoods $\mathcal{L}_k$ for each simulated observation $k$ using a $\chi^2$ comparison:

\begin{equation}
\mathcal{L}_k(\Delta m_\rmn{micro}|ln(r_s), p, s) = \rmn{exp}\left(\frac{-\chi_k^2}{2}\right).
\end{equation}
The final likelihood for a given combination of parameters $ln(r_s)$, $p$, $s$ is simply the sum over all of the likelihoods of the individual simulated observations $\mathcal{L} = \sum_k \mathcal{L}_k$.

Differential probability distributions are generated from the likelihoods using Bayes' theorem:

\begin{equation}
\frac{\rmn{d}^3P}{\rmn{d}ln(r_s)\rmn{d}p\rmn{d}s} \propto \mathcal{L}(\Delta m_\rmn{micro}|ln(r_s), p, s)\frac{\rmn{d}P_\rmn{prior}}{\rmn{d}ln(r_s)}\frac{\rmn{d}P_\rmn{prior}}{\rmn{d}p}\frac{\rmn{d}P_\rmn{prior}}{\rmn{d}s}.
\end{equation}
We use flat priors for all three parameters (equivalent to a logarithmic prior on the source size $r_s$). Since we are interested in the quasar accretion discs, we integrate over $\rmn{d}s$ as a nuisance parameter. Results are presented as mean values with 68 per cent confidence intervals, or 68 per cent upper/lower limits where appropriate.

\section{Results}
\label{sec:results}

\begin{figure*}
\begin{center}
  \subfigure[\mg]{
    \label{fig:0414}	
    \includegraphics[width=70mm]{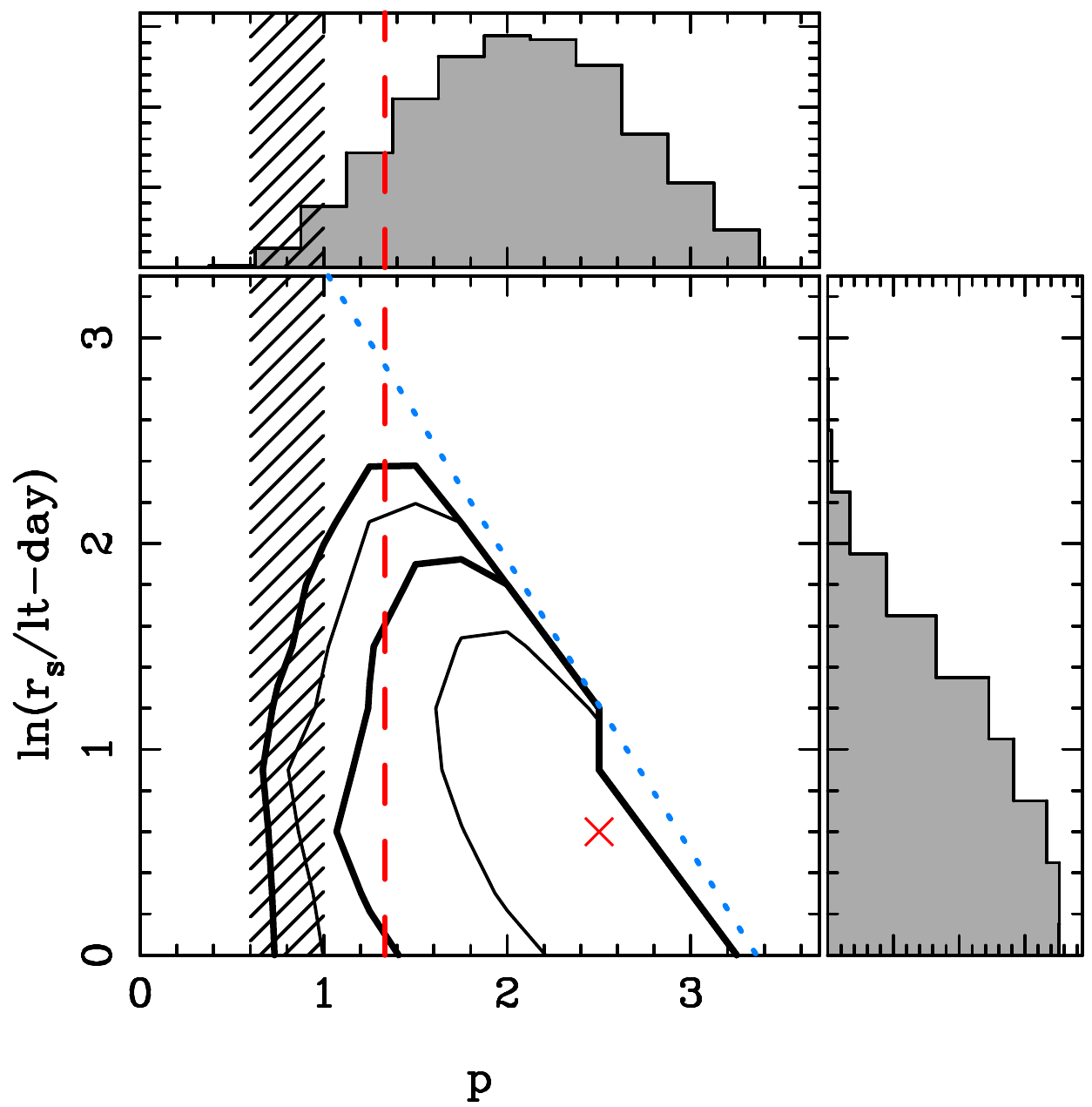}
  }
  \subfigure[\rxj]{
    \label{fig:0911}	
    \includegraphics[width=70mm]{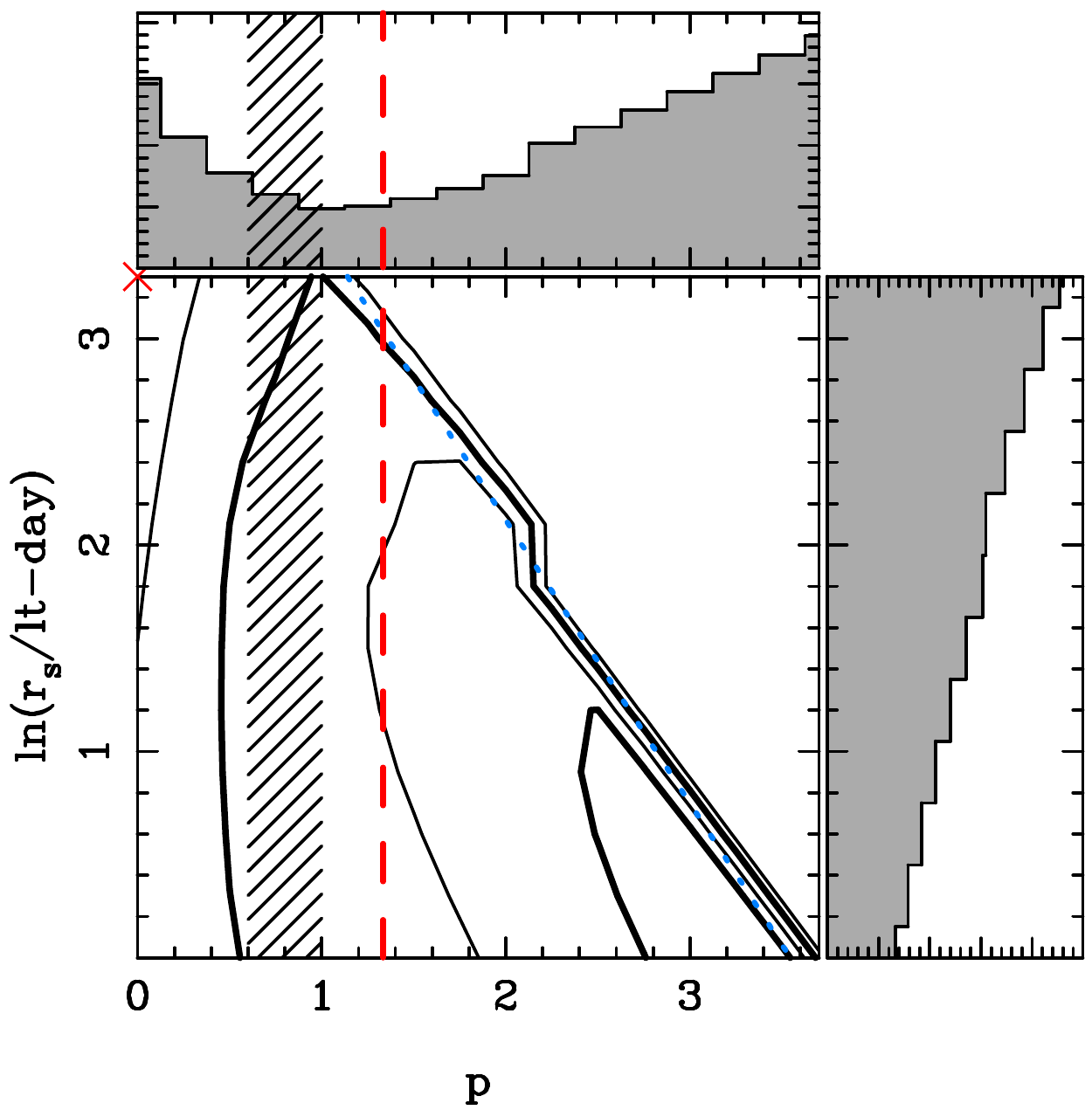}
  }\\
  \subfigure[\qsob]{
    \label{fig:1422}
    \includegraphics[width=70mm]{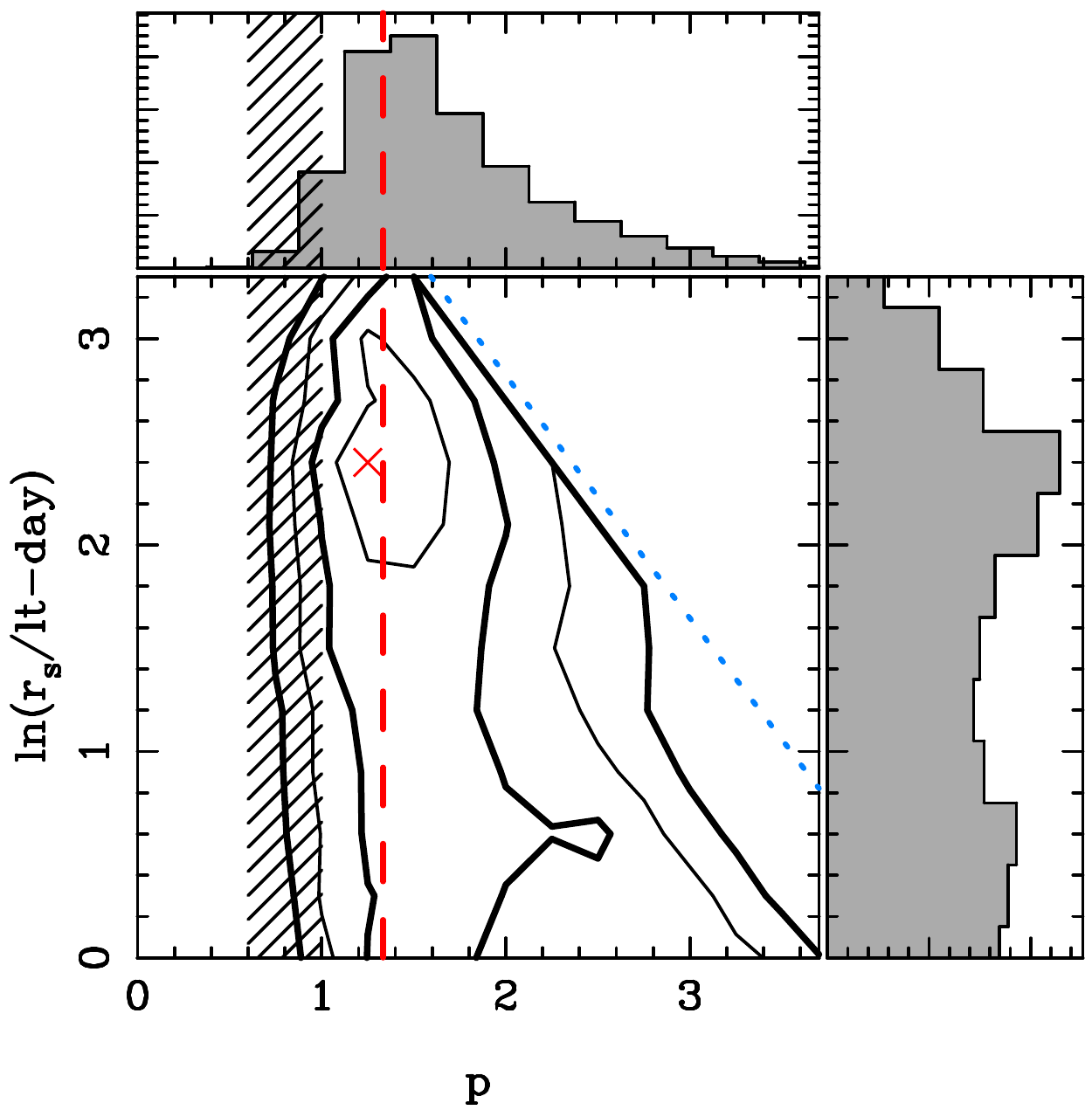}
  }
  \subfigure[\wfi]{
    \label{fig:2026}
    \includegraphics[width=70mm]{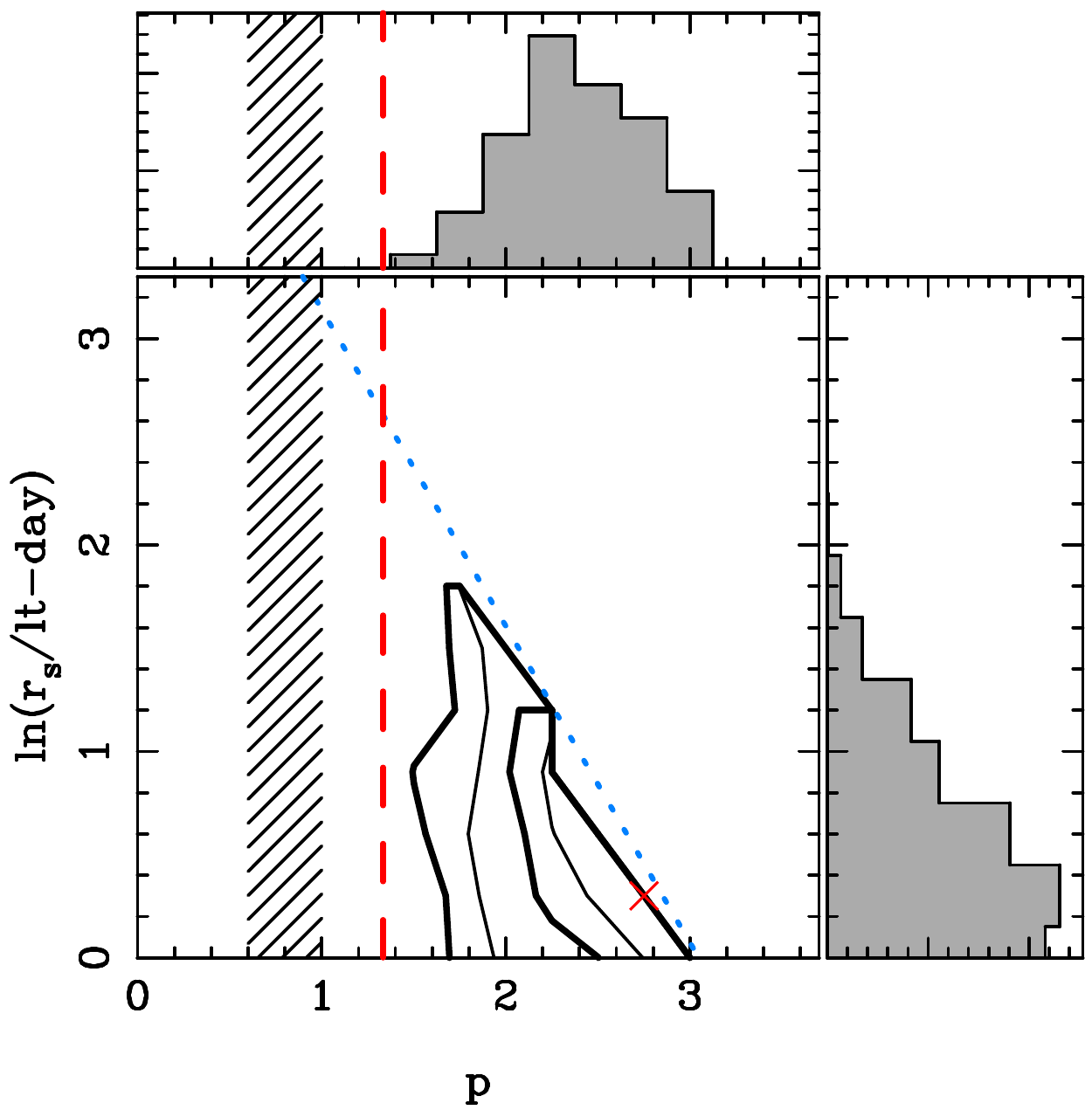}
  }
  \end{center}
  \caption{Accretion disc constraints for four lensed quasars using our HST data only. Contours are drawn at intervals of $0.5\sigma$ (thick lines at $1\sigma$ and $2\sigma$) for two degrees of freedom from the maximum. The solid contours therefore formally contain 68.3 and 95.4 per cent of the probability. The dashed red line marks the power-law index for the SS disc \citep{ss73}. The hatched region is the joint $1\sigma$ constraint from JV14. The red cross marks the peak probability in the two-dimensional surface, and the dotted light blue line is the limit of our simulations (see Section \ref{sec:results}). Panel (a): \mg. Panel (b): \rxj. Panel (c): \qsob. Panel (d): \wfi.}
  \label{fig:constraints}
\end{figure*}

Probability distributions for the accretion disc parameters in each of our four systems are provided in Figure \ref{fig:constraints}, marginalised over the smooth matter fraction and errors in the unmicrolensed baseline. In each figure we have marked the power law index of an SS disc with a dashed red line ($p = 4/3$), and the Bayesian ensemble estimate from JV14 with a hashed region ($p = 0.8 \pm 0.2$).

The dotted light blue line in each figure marks the point at which the combination of parameters $ln(r_s)$ and $p$ gives a source size in the reddest filter that is larger than we can fit in our 100 Einstein radius maps (this limit is a Gaussian dispersion of 8 Einstein radii). Beyond this point, we simply assume that the simulated flux ratio equals the unmicrolensed macro-flux ratio. In most cases this is a reasonable assumption, however large-scale structures in sheared magnification maps (conglomerations of caustics stacked on top of each other) can still cause slight deviation from the macro-flux ratio, even at these sizes.

Although there may be some probability that the true accretion disc parameters lie beyond this dotted light blue line (specifically in \mg~ and \wfi), we do not expect it to be significant. In any case, if we expanded the maximum size limitations in our simulations the measured $p$ constraints could only skew to higher values, pushing them further away from the JV14 result.

The formal 68 per cent constraints obtained from these probability distributions are provided in Table \ref{tab:constraints}. We quote results after marginalising over the smooth matter fraction $s$, as well as the constraints obtained if we assume $s=0.8$. The most notable feature of these measurements and the probability distributions in Figure \ref{fig:constraints} is their diversity, particularly when compared with JV14 or \citet{blackburne+11}. The reasons for these differences will be discussed in the following section.

\begin{table*}
\centering
  \caption{Measured accretion disc constraints (68 per cent)} \label{tab:constraints}
  \begin{tabular}{l|cc|cc} 
  \hline
  & \multicolumn{2}{c}{Marginalised} & \multicolumn{2}{c}{80 per cent smooth matter} \\
 System & $r_s$ (lt-day) & $p$ & $r_s$ (lt-day) & $p$ \\
 \hline
 \mg & $< 2.9$ & $2.1^{+0.6}_{-0.6}$ & $< 2.5$ & $1.8^{+0.6}_{-0.6}$ \\ 
 \rxj & -- & -- & -- & -- \\
 \qsob & $5.1^{+8.2}_{-3.5}$ & $1.6^{+0.7}_{-0.4}$ & $6.8^{+5.6}_{-4.4}$ & $1.4^{+0.5}_{-0.4}$ \\
 \wfi & $<2.2$ & $2.4^{+0.4}_{-0.4}$ & $<2.9$ & $2.3^{+0.5}_{-0.4}$ \\
 \hline
\end{tabular}
\end{table*}

\begin{figure*}
  \includegraphics[width=125mm]{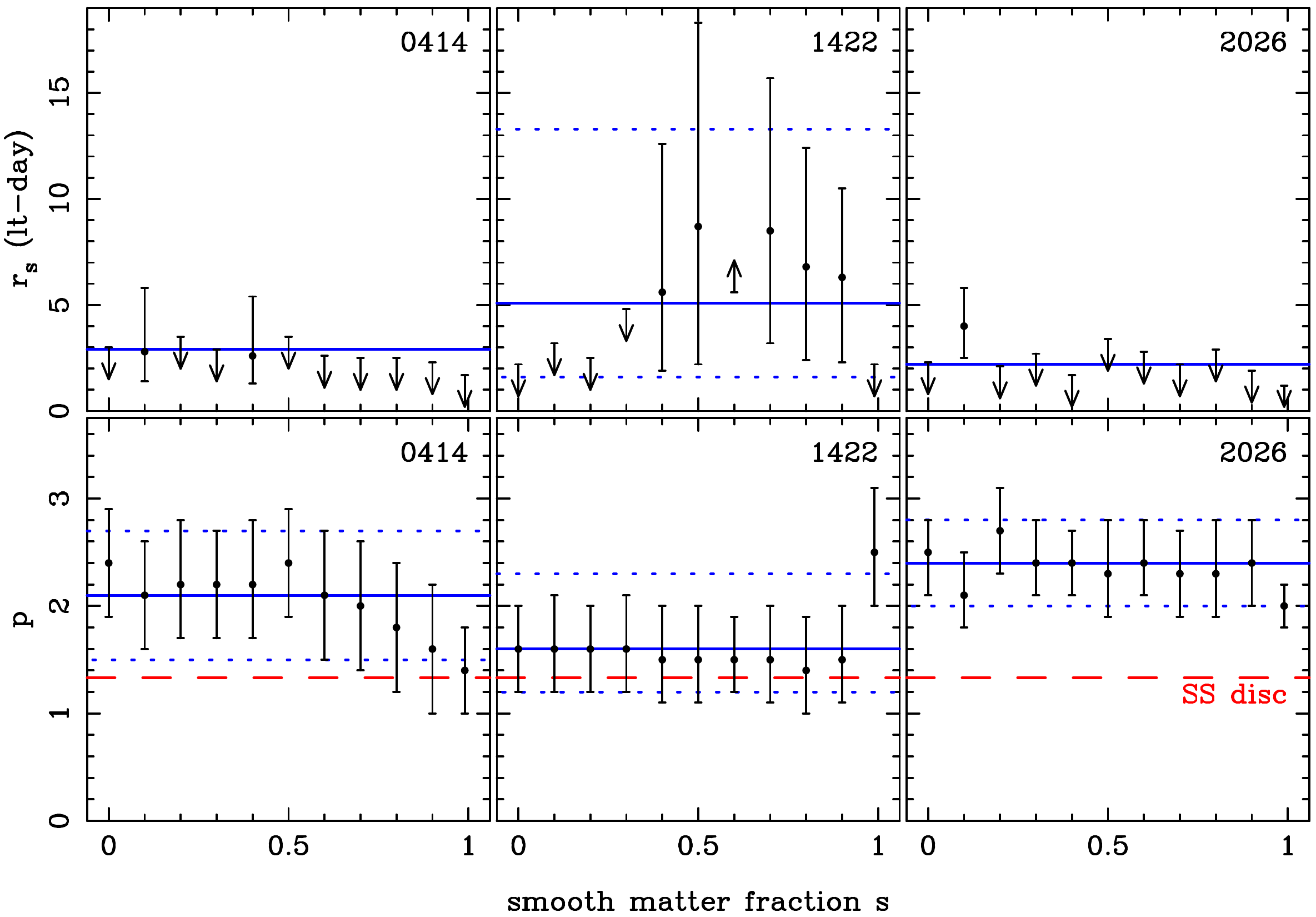}
  \caption{Variation of accretion disc constraints as a function of smooth matter fraction $s$. Error bars denote 68 per cent confidence. Arrows are 68 per cent upper limits ($s=0.6$ in \qsob~is a lower limit). The top panels show size constraints $r_s$. The bottom panels show constraints on the temperature profile power-law index $p$. Blue lines indicate the constraints obtained when marginalising over $s$. In the top left (\mg) and top right (\wfi) panels, these are 68 per cent upper limits. In all other panels, the solid blue lines show the marginalised constraint, and the dotted blue lines the 68 per cent confidence limits. The Shakura-Sunyaev power-law index $p=4/3$ is plotted as a dashed red line in the bottom panels.}
  \label{fig:smooth}
\end{figure*}

We show the explicit dependence of our accretion disc constraints on the assumed smooth matter fraction $s$ in Figure \ref{fig:smooth} (\rxj~is excluded from this figure; the data for this system do not provide constraints at any smooth matter fraction). With a few exceptions, we obtain only upper limits on the size $r_s$ of the accretion disc at $\lambda_0 = 1026$\AA~(in \qsob~we obtain a measurement, but its error bars are large). For the power-law index $p$ we find the marginalised constraints are completely consistent with each of the individual smooth matter cases, tending to deviate only at $s=0.99$. There is a slight trend towards decreasing $p$ (steepening temperature profile) with increasing $s$, most pronounced in \mg.

For the remainder of this paper, we restrict ourselves to discussing results for the $s=0.8$ case. We choose this value following \citet{jv+15a}, who measured $s = 0.79\pm0.14$ from a sample of 19 lensed quasars. We note, however, that this specific constraint is not uniformly applied in the literature: for example, \citet{rojas+14} and \citet{motta+17} use $s=0.9$, and JV14 used $s=0.95$.

\section{Discussion}
\label{sec:discuss}

\subsection{Accretion disc size $r_s$}
The accretion disc sizes quoted in Table \ref{tab:constraints} are Gaussian dispersions, assuming $1M_\odot$ microlenses. To facilitate comparisons with thin-disc theory, we convert to half-light radii $r_{1/2} = 1.18r_s$. The average stellar microlens in a lensing galaxy is likely to be less massive than $1M_\odot$; the standard value used is $M_\rmn{micro}=0.3M_\odot$, which enters into our source sizes as a factor of $(M_\rmn{micro}/M_\odot)^{1/2}$. Our resulting half-light radius measurements assuming $0.3M_\odot$ microlenses are provided in Table \ref{tab:sizes}.

\begin{table*}
\centering
  \caption{Measured accretion disc half-light radii (68 per cent)} \label{tab:sizes}
  \begin{tabular}{lccc} 
  \hline
 System & $r_{1/2}$ (lt-day) & $M_\rmn{BH}$ ($10^9M_{\odot}$)$^a$ & $r_{1/2} / r_{\rmn{thin}}$ \\
 \hline
  \mg & $<1.6$ & 1.82 & $<1.6$ \\ 
 \rxj & -- & 0.80 & -- \\
 \qsob & $4.4^{+3.6}_{-2.8}$ & 4.79 & $2.3^{+1.9}_{-1.5}$ \\
 \wfi & $<1.9$ & 1.00$^b$ & $<2.8$ \\
 \hline
  \multicolumn{4}{l}{Results assuming smooth matter fraction $s=0.8$}\\
  \multicolumn{4}{l}{$^a$ From \citet{mosquera+11}}\\
  \multicolumn{4}{l}{$^b$ Fiducial value; no measured black hole mass available}\\  
\end{tabular}
\end{table*}

The half-light radius of a standard thin disc \citep{ss73} is:
\begin{equation}
\label{eqn:thindisc}
\begin{aligned}
r_{\rmn{thin}} &= 2.44\left(\frac{45G^2M^2_\rmn{BH}m_pf_\rmn{Edd}\lambda^4}{4\pi^5h_pc^3\sigma_T\eta}\right)^{1/3}\sqrt{\rmn{cos}\ i}\\
& =6.49 \rmn{ld}  \left(\frac{M_\rmn{BH}}{10^9M_\odot}\right)^{2/3} \left(\frac{f_\rmn{Edd}}{\eta}\right)^{1/3} \left(\frac{\lambda}{\mu\rmn{m}}\right)^{4/3},
\end{aligned}
\end{equation}
where $M_\rmn{BH}$ is the black hole mass, $m_p$ is the proton mass, $h_p$ is Planck's constant, $\sigma_T$ is the Thomson scattering cross-section, and $i$ is the inclination angle (assumed to be $\langle\rmn{cos}\ i\rangle = 1/2$). $f_\rmn{Edd}$ is the Eddington ratio, the ratio of the quasar's bolometric luminosity to the Eddington luminosity, and $\eta$ is the accretion efficiency. Finally, $\lambda$ is the rest wavelength of interest.

Black hole mass measurements are available for three of our systems: \mg, \rxj, and \qsob~\citep{mosquera+11}. For \wfi~we assume a fiducial $M_\rmn{BH} = 10^9 M_\odot$. In Table \ref{tab:sizes} we provide black hole masses and ratios of measured half-light radius $r_{1/2}$ to the thin disc prediction $r_{\rmn{thin}}$, calculated using Equation \ref{eqn:thindisc}.

For \mg~and \wfi, our two systems showing the largest chromatic variation, we obtain only upper limits. Nevertheless, in line with previous microlensing analyses, we find that our measured sizes are consistent with being larger than expected from thin disc theory (only marginally, in the case of \mg). In JV14, they quoted an ensemble measurement of $r_s = 4.5^{+1.5}_{-1.2}$ light days for eight quasars (assuming $1M_\odot$ microlenses). Converting that to half-light radius as described above, their constraint is $r_{1/2} = 2.9^{+1.0}_{-0.8}$ light days. This is marginally larger than our measurement for \mg, and completely consistent with our \qsob~and \wfi~results. 
\\
\subsection{Power-law index $p$}
In JV14, the authors report a Bayesian estimate of the power-law index $p$ of $0.8\pm0.2$, measured jointly from 8 lensed quasars. The accretion disc temperature profile goes as $T \propto r^{-1/p}$, so this corresponds to a steeper temperature profile than expected for an SS disc.

This steep temperature profile measurement is difficult to understand in the context of microlensing light curve constraints on accretion disc sizes (e.g. \citealt{morgan+10}), or recent reverberation mapping experiments (e.g. \citealt{starkey+17}; \citealt{jiang+17}), which consistently measure accretion discs to be factors of $\sim4$ larger than expected from thin disc theory. Naively, you would expect that larger accretion discs at a given wavelength imply shallower temperature profiles.

We have measured $p$ in four additional systems, and we prefer \textit{larger} values of $p$ than $4/3$ in two of our systems: \wfi~and \mg~(marginally). Both of these results are consistent with the $p=2.0$ slim disc prescription of \citet{abramowicz+88}. In the third system, \qsob, we find $p\sim4/3$ is preferred, consistent with standard SS thin discs. Our \rxj~data are insufficient to place any constraints on its accretion disc structure.

\begin{figure}
  \includegraphics[width=85mm]{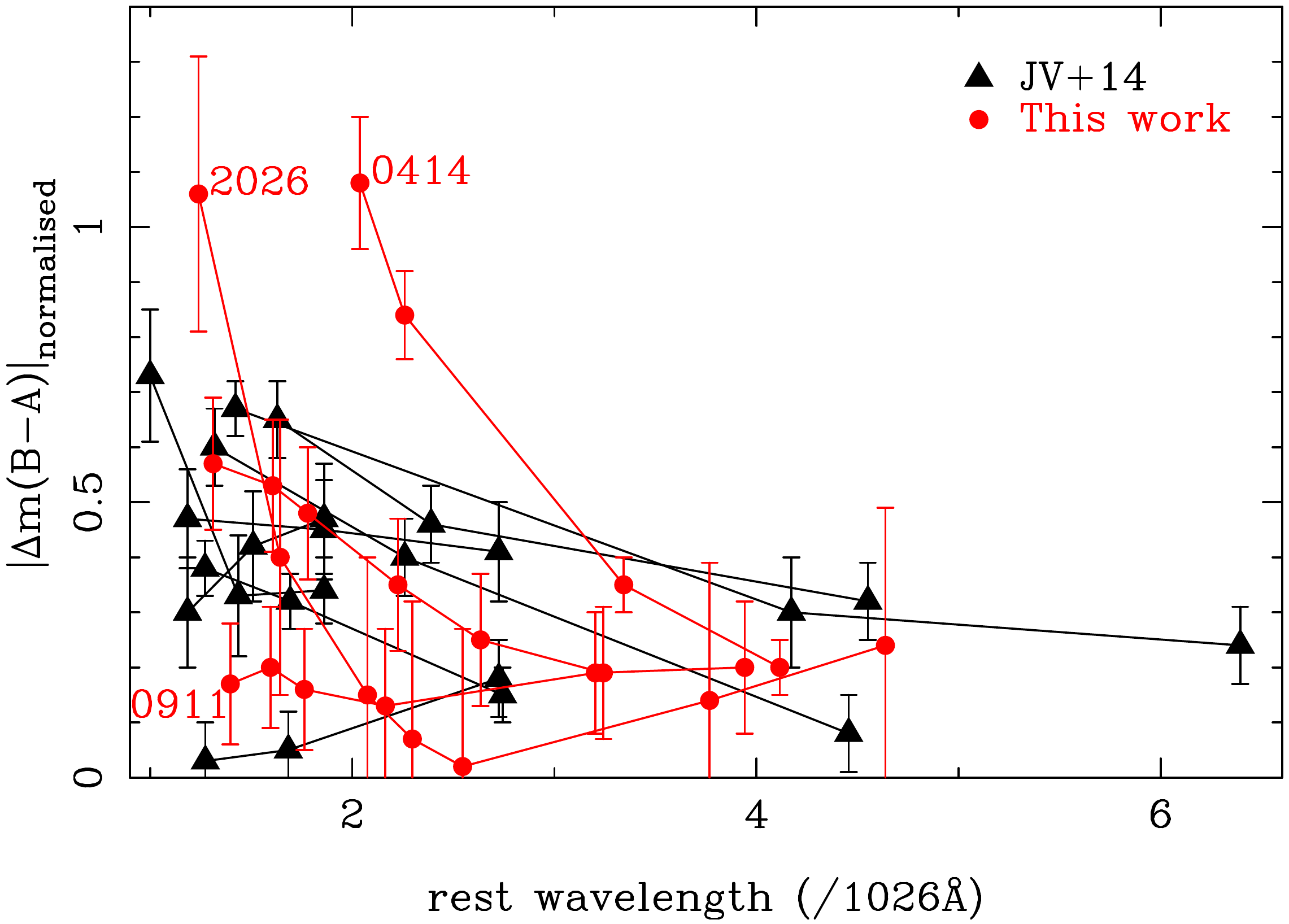}
  \caption{Absolute values of observed magnitude differences between lensed images, normalised to their expected macro-magnifications (errors in our HST data are largely dominated by errors in the expected macro-magnifications; see Tables \ref{tab:ratios} and \ref{tab:dm_micro}). The red circles are four sets of observations reported in this paper, whereas the black triangles are the sample from JV14 (excluding Q 2237+0305, the data for which were not provided in that paper). $B-A$ is adopted for labelling convenience; in \mg~and \wfi, $A = A_1$ and $B = A_2$. The label for \qsob~is omitted from the plot for clarity. It is clear from this figure that \mg~and \wfi~show a much greater degree of chromatic variation than any of the systems in the JV14 sample.}
  \label{fig:obs}
\end{figure}

Why do we find such a variety of constraints on $p$, where the measurements in JV14 were largely consistent with each other, converging on $p=0.8\pm0.2$? In Figure \ref{fig:obs}, we plot all of the observations reported in JV14 (black triangles, from their Table 1) and this paper (red circles). The figure shows absolute magnitude differences between lensed images as a function of rest wavelength in each quasar. The variation in absolute magnitude difference between shortest and longest wavelength represents the degree of microlensing-induced chromatic variation in each system.

Figure \ref{fig:obs} clearly illustrates that two of our systems -- \mg~and \wfi~ -- display more chromatic variation than any of the objects studied in JV14. In contrast, \qsob~shows chromatic microlensing effects broadly similar to the JV14 sample, and \rxj~has essentially no chromatic microlensing at all. Note that there is no overlap between the systems studied here, and those in JV14.

\begin{figure}
  \includegraphics[width=85mm]{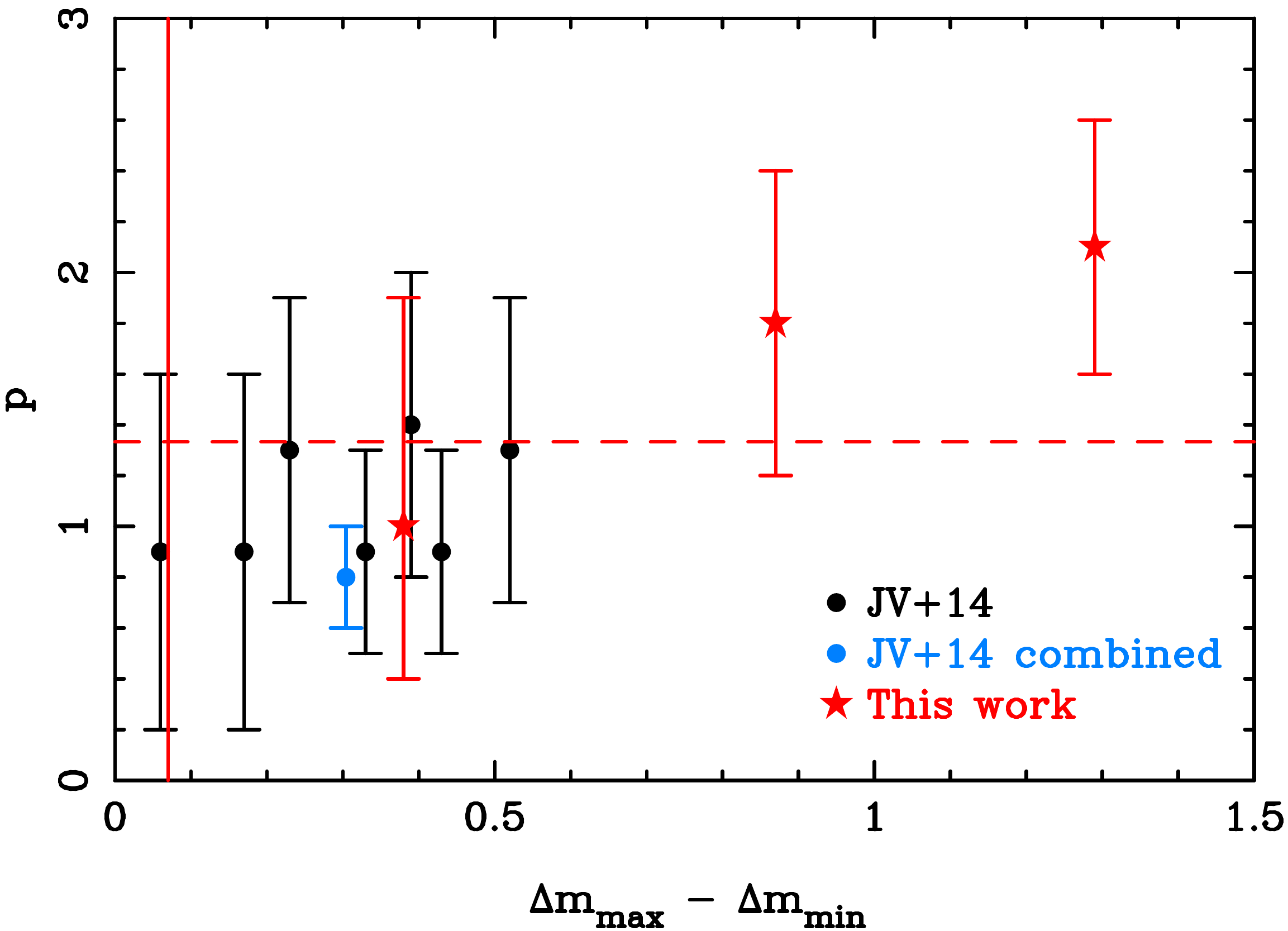}
  \caption{Power-law constraints $p$ ($68$ per cent confidence) as a function of the size of chromatic variation in the observation. As the degree of chromatic variation increases, so does the measured constraint on the power-law index. Seven quasars are presented from \citet{jv+14} (filled black circles, Bayesian constraints), along with their joint constraint at the average chromatic variation of their sample (light blue filled circle). Three quasars from this work are shown (red stars, $s=0.80$), with a vertical red line marking the degree of chromatic variation in the fourth (\rxj, where we obtain no constraint). The red dashed line marks the power-law index $p=4/3$, appropriate for the standard SS disc.}
  \label{fig:chromaticity}
\end{figure}

In cases where we observe strong chromatic effects, we find $p>4/3$, whereas in cases where chromatic effects are weaker we find $p$ consistent with (or even smaller than) $4/3$. If we look at the individual quasars in the JV14 sample, two of the three systems displaying the strongest chromatic effects also provide the largest estimates of the power-law index (HE 0512-3329: maximum likelihood $p = 1.25^{+0.6}_{-0.7}$, Bayesian $p=1.4\pm0.6$; SDSS J1004+4112: maximum likelihood $p = 1.00^{+1.00}_{-0.5}$, Bayesian $p=1.3\pm0.6$). This trend is illustrated in Figure \ref{fig:chromaticity}, where we plot the observational constraints for each system in JV14 (black) and this work (red) as a function of the size of their chromatic variation. The latter is defined as $\Delta m_\rmn{max} - \Delta m_\rmn{min}$ (usually, but not always, the bluest observation minus the reddest). The error bars in the individual observations are large; formally, they are all consistent with each other, and with the SS disc prediction. Nevertheless, there does appear to be a general trend, as described above: observations with lower chromatic variation tend to predict lower values of $p$.

Assuming this trend is real, is it due to the physical situation in each individual quasar? Do the quasars where we observe lower chromatic variation actually have steeper temperature profiles in their accretion discs? Or is it simply a (misleading) selection effect, where observations with low chromatic variation lead us to believe we are observing a steeper temperature profile than actually exists in the quasar? We can check this with mock simulations, where we test how well we are able to recover a known input accretion disc using our single-epoch microlensing analysis technique. We analyse a sample suite of mock simulations below.

\section{Mock observations}
\label{sec:mocks}

To explore whether the correlation between size of the observed chromatic variation and the slope of the measured temperature profile is real or simply a selection effect, we generated a suite of 200 mock observations for a single input accretion disc. These were conducted in a blind fashion -- one of us (GV) produced the mock data, and provided it as a set of magnitude differences to another (NFB), who ran them through the microlensing analysis pipeline. The purpose of these mock observations was twofold: 1) to test whether observations with low chromatic variation produce accretion disc constraints biased towards low $p$, and 2) to test whether any observations recover the input disc parameters. 

\subsection{Mock observation technique}

The same machinery was used to produce the mock observations as described in Section \ref{subsec:technique}. We used Equation \ref{eqn:profile} to create a series of Gaussian profiles to describe the shape of the accretion disc for each observed wavelength in Table \ref{tab:ratios}, with the parameters $p=1.5$ and $r_s=2.7$ light days. We did this using the \mg~macromodel, and the magnification maps we have already generated. For simplicity, we assumed the magnification for image $A_1$ to be constant and equal to the macro-magnification. We then selected a map with a fixed $s$ for image $A_2$, convolved it with the generated source profiles, and sampled magnification values from the central $6700 \times 6700$ pixel area. This area was chosen in order to avoid convolution edge effects; the largest source profile, for observed wavelength $\lambda = 15369$\AA, corresponds to roughly 3300 pixels in the source plane. 

We used $s=0.8$ and $s=0.9$ magnification maps to generate the mock data, drawing 100 mock observations from each map. These smooth matter fractions are consistent with the values we expect to find in a real \mg-like quasar. Using two smooth matter fractions allowed us to generate more statistically-independent mock observations, and also to test whether the choice of $s$ had an effect on our results. Unsurprisingly, we found that it did not -- the two cases were indistinguishable.

Throughout, we have assumed errors on our mock observations of $\pm0.05$ in the UVIS channel, and $\pm0.02$ in the IR channel. These are consistent with our HST observations. Although simulated magnitude differences were generated for nine HST filters (see Table \ref{obsdetail}), we used only five for the microlensing analysis: F410M ($4109$\AA), F621M ($6219$\AA), F845M ($8436$\AA), F125W ($12486$\AA), and F160W ($15369$\AA). 

\subsection{Mock observation results}
\label{subsec:mock_results}

\begin{figure*}
  \includegraphics[width=120mm]{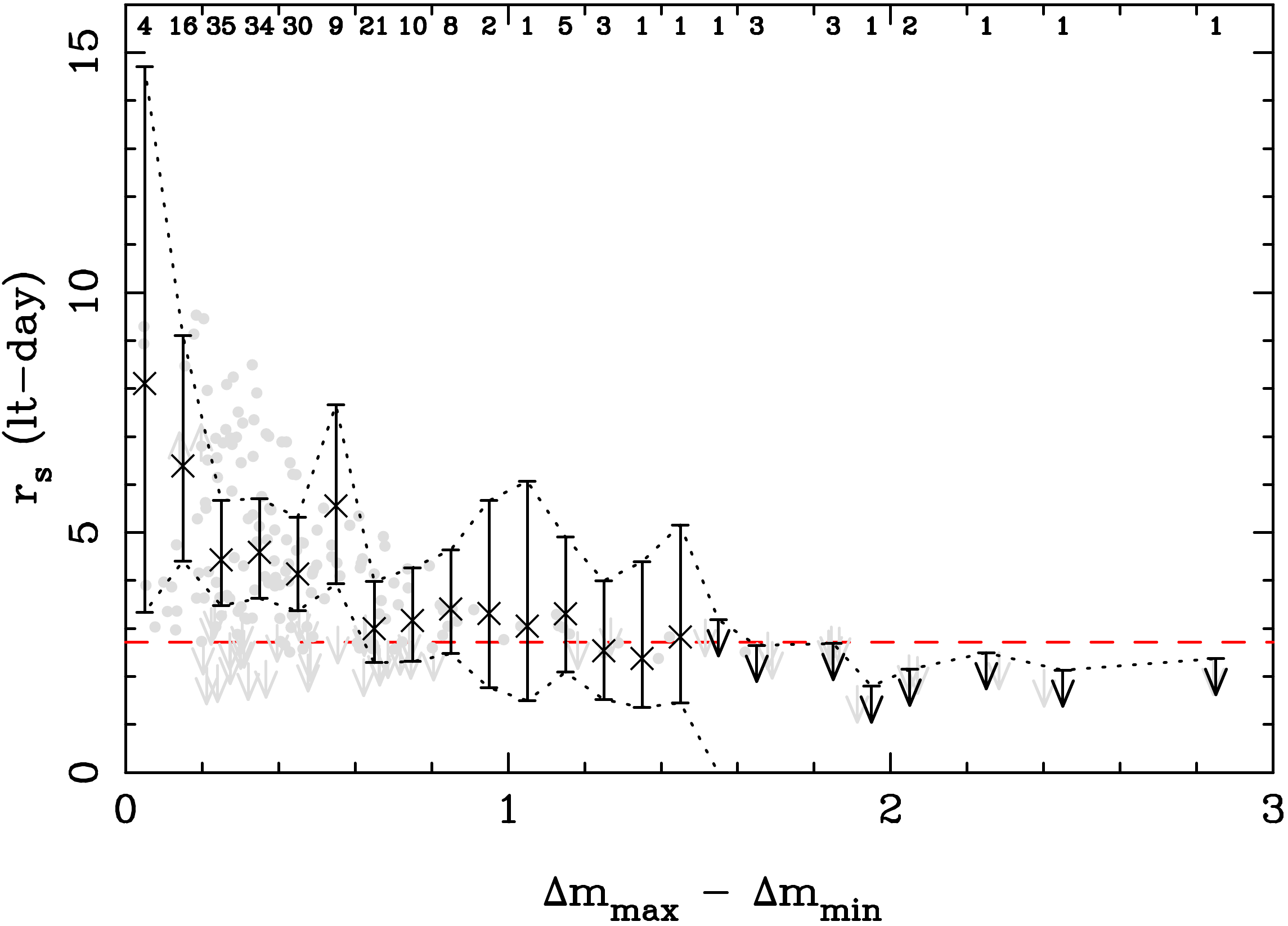}
  \caption{Quasar accretion disc size constraints $r_s$ as a function of the size of chromatic variation in mock observations. Grey points indicate individual measurements for 193 mock observations (error bars excluded for clarity). The dashed red line is the input value of $r_s=2.7$ light days. Black crosses are the result of combining all likelihood surfaces in chromatic variation bins with a width of 0.1 (the number of mock observations in each bin is shown across the top of the figure). All error bars are 68 per cent confidence intervals; arrows indicate 68 per cent upper limits.}
  \label{fig:mock_rs}
\end{figure*}

\begin{figure*}
  \includegraphics[width=120mm]{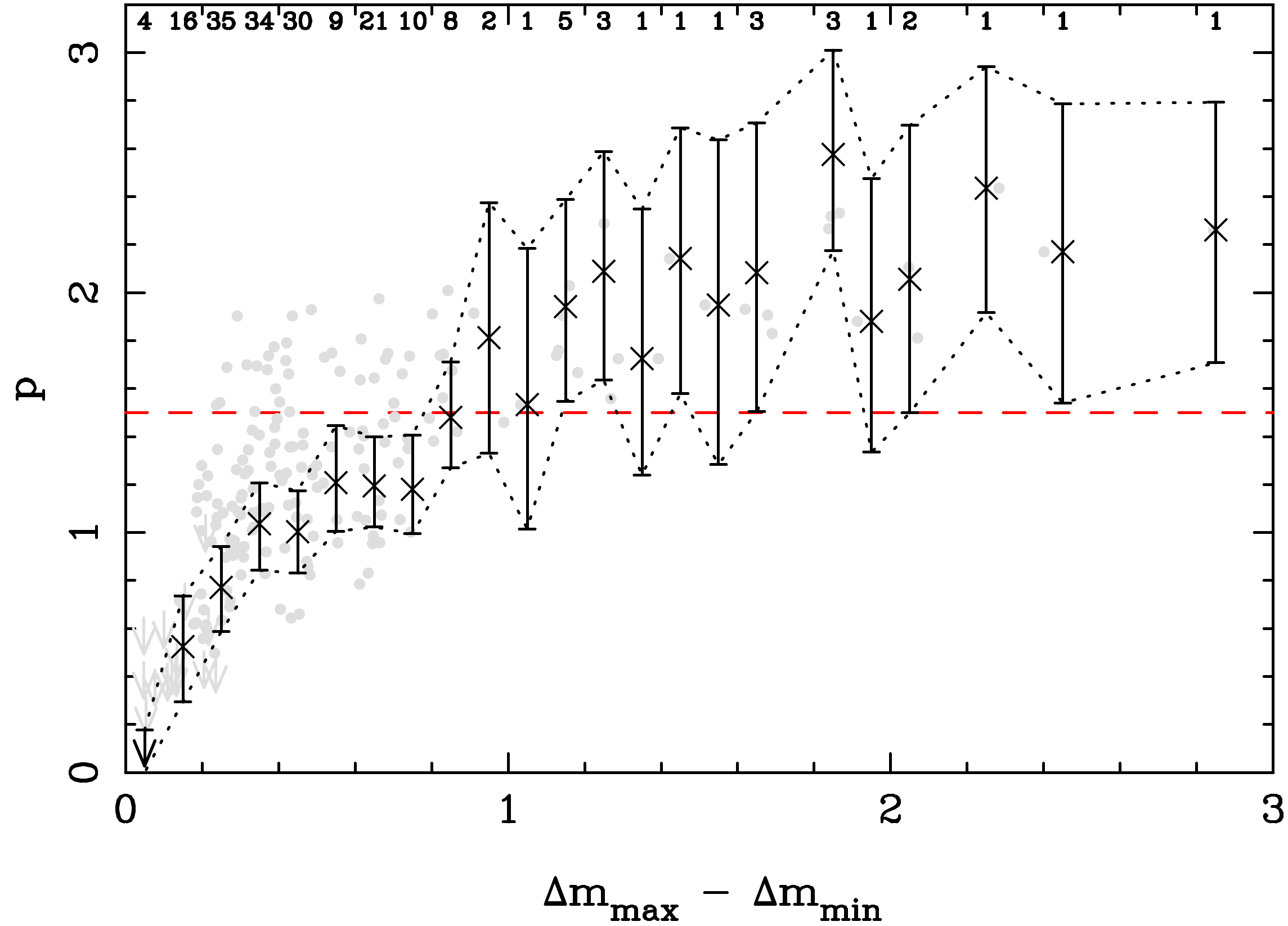}
  \caption{Power-law temperature profile constraints $p$ as a function of the size of chromatic variation in mock observations. Grey points indicate individual measurements for 193 mock observations (error bars excluded for clarity). The dashed red line is the input value of $p=1.5$. Black crosses are the result of combining all likelihood surfaces in chromatic variation bins with a width of 0.1 (the number of mock observations in each bin is shown across the top of the figure). All error bars are 68 per cent confidence intervals; arrows indicate 68 per cent upper limits.}
  \label{fig:mock_p}
\end{figure*}

The results of single-epoch microlensing analysis on our mock observations are presented in Figures \ref{fig:mock_rs} (source size $r_s$) and \ref{fig:mock_p} (power-law index $p$), plotted as a function of chromatic variation $\Delta m_\rmn{max} - \Delta m_\rmn{min}$. We have chosen to plot only the simulation results where we assumed a smooth matter fraction of $s=0.8$ in the analysis pipeline; we also ran the microlensing simulations marginalising over all smooth matter fractions, and found all of the same trends.

The grey points in Figures \ref{fig:mock_rs} and \ref{fig:mock_p} show the individual constraints for each of the mock simulations. The error bars on the individual simulations have been excluded for clarity. Errors on any single measurement are broad, corresponding roughly to the sizes of the observed constraints in Figure \ref{fig:chromaticity}. A total of 193 points are plotted: seven mock observations returned no meaningful constraints.

We have also combined the likelihood surfaces in bins of $\Delta m_\rmn{max} - \Delta m_\rmn{min} = 0.1$, treating each mock observation as an independent observation of the same source. These combined results are plotted as black crosses with error bars in Figures \ref{fig:mock_rs} and \ref{fig:mock_p}; across the top of each plot, we show the number of mock observations in each bin. Red dashed lines show the input parameters of $r_s = 2.7$ light days and $p = 1.5$.

These illustrative mock simulations demonstrate exactly the trend seen in the observational data: smaller chromatic variation leads to smaller estimated power-law indices $p$. They also result in larger estimated source sizes $r_s$. We note that the simulations presented in \citet{jv+14} cover a range in chromatic variation of roughly $0.04 < \Delta m_\rmn{max} - \Delta m_\rmn{min} < 0.52$. The four quasars presented in this paper span the much larger range from $0.07$ (\rxj) to $1.29$ (\wfi).

Mock observations with chromatic variations larger than $\Delta m_\rmn{max} - \Delta m_\rmn{min} = 1.0$ are rare; in most bins we have only one or a few simulated data points. There is some suggestion that we undershoot (overshoot) the input $r_s$ ($p$) for sufficiently large chromatic variation, however more mock observations are required to robustly explore this part of parameter space.

\begin{figure*}
  \includegraphics[width=120mm]{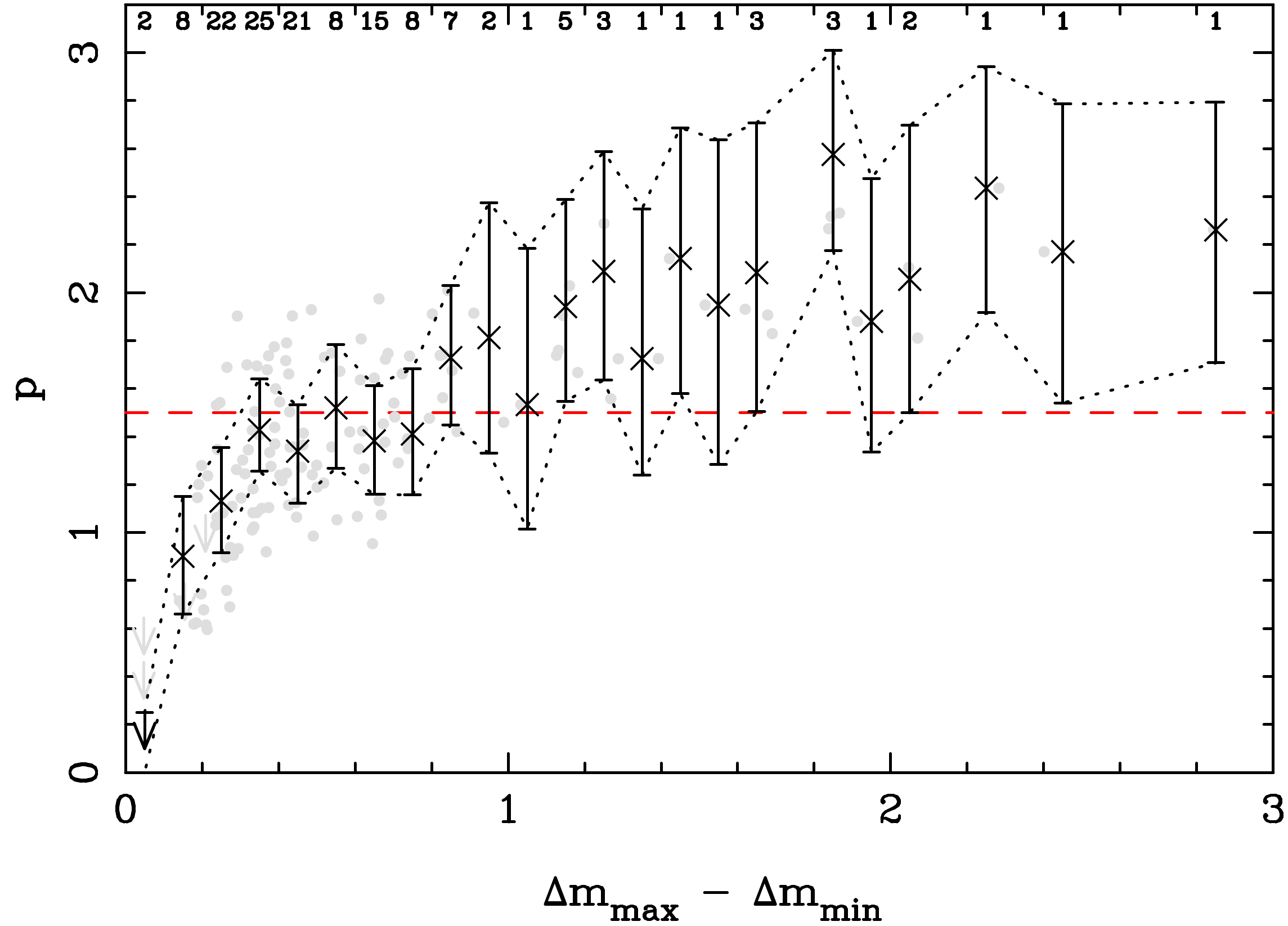}
  \caption{Per Figure \ref{fig:mock_p}, using only mock observations with $|\Delta m_\rmn{red}-\Delta m_\rmn{macro}| < 0.3$. This leaves 110 mock observations, all of which converge to the unmicrolensed baseline within the applied limit. This cut results in improved recovery of the input power-law index $p$ in the range $\Delta m_\rmn{max} - \Delta m_\rmn{min} < 1.0$.}
  \label{fig:mock_offset}
\end{figure*}

Nevertheless, it does appear that for chromatic variation above a certain threshold ($\Delta m_\rmn{max} - \Delta m_\rmn{min} \approx 0.6$) we do recover the input simulation parameters relatively well. We find that we can improve our recovery rate by using only mock observations that roughly converge to the theoretically-expected macro-magnification in their reddest filter. Using a convergence criterion of $\Delta m_\rmn{red} - \Delta m_\rmn{macro} < 0.3$, we are able to push the threshold of useful data down to chromatic variations of $\Delta m_\rmn{max} - \Delta m_\rmn{min} \approx 0.4$. In Figure \ref{fig:mock_offset} we plot the 110 mock observations that meet this convergence limit.

Figure \ref{fig:mock_offset} clearly shows that in our suite of 200 mock observations, we are able to reliably recover the input value of $p=1.5$ in ensemble measurements if we restrict ourselves to observations with chromatic variation of $\Delta m_\rmn{max} - \Delta m_\rmn{min} > 0.4$, and convergence to the macro-magnification of $\Delta m_\rmn{red} - \Delta m_\rmn{macro} < 0.3$.

\subsection{Ensemble measurements}

\begin{figure*}
\begin{center}
  \subfigure[Mock combined (JV14-like)]{
    \label{fig:mock_jv}	
    \includegraphics[width=55mm]{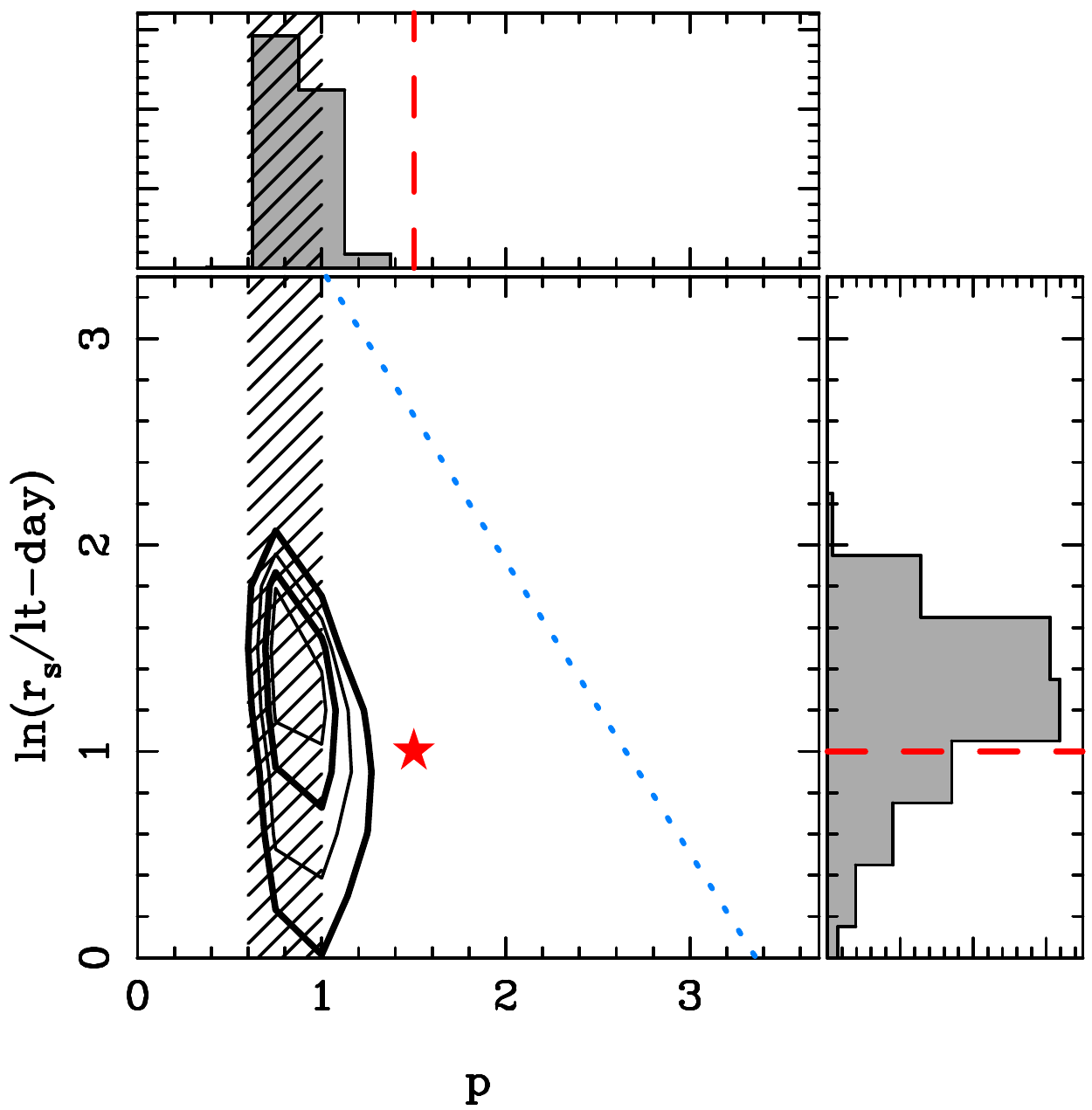}
  }
  \subfigure[Mock combined (random)]{
    \label{fig:mock_all}	
    \includegraphics[width=55mm]{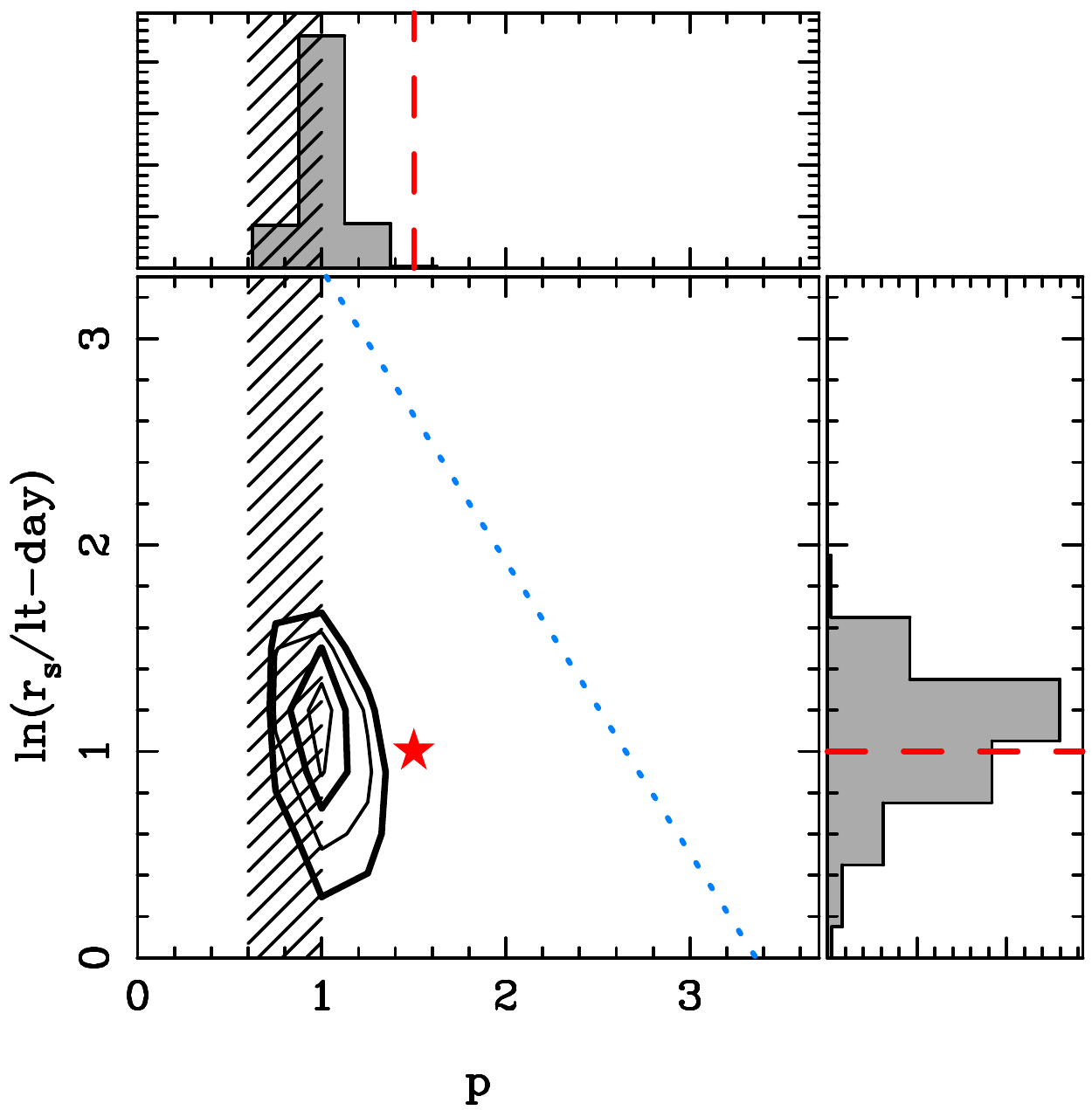}
  }
  \subfigure[Mock combined (constrained)]{
    \label{fig:mock_good}	
    \includegraphics[width=55mm]{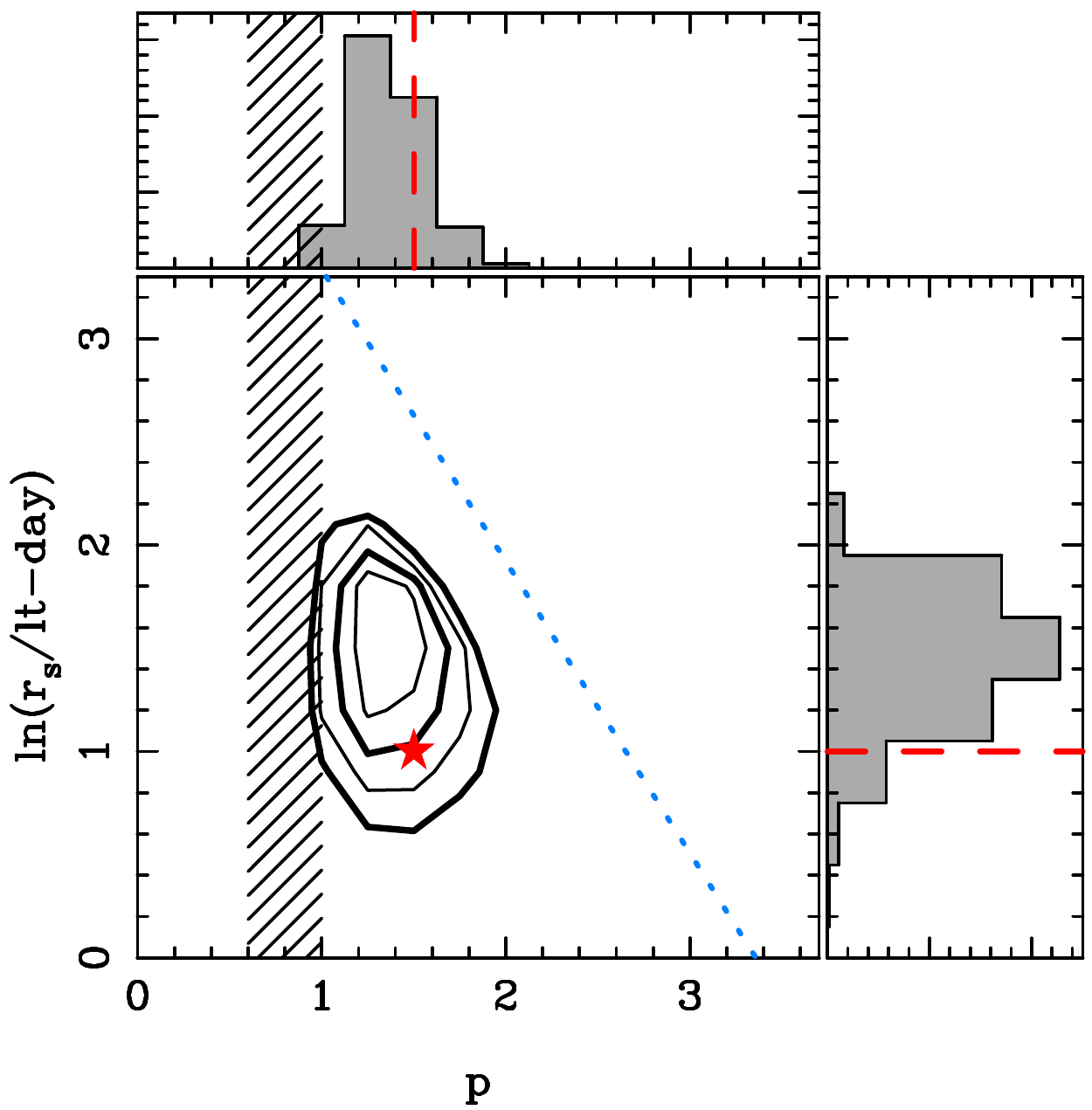}
  }  
  \end{center}
  \caption{Illustrative constraints obtained by combining eight randomly-chosen mock observations of an accretion disc with $p = 1.5$, $ln(r_s/\text{lt-day})=1.0$ (red star and dashed red lines), as per Figure \ref{fig:constraints}. Panel (a) shows the results when combining observations constrained to match the range of chromatic variation seen in the JV14 data. Panel (b) shows the combination of observations chosen without restriction from our mock data. In Panel (c) we combine mock observations using constraints on chromatic variation and convergence to the unmicrolensed baseline described in Section \ref{subsec:mock_results}.}
  \label{fig:mock_constraints}
\end{figure*}

These mock simulations imply that using observations with low chromatic variation leads to systematically underestimating the temperature profile power-law index $p$. This bias towards smaller $p$ in low chromatic variation observations can be particularly misleading when combining data. We illustrate this in Figure \ref{fig:mock_constraints}. Here, we have randomly selected eight mock observations and combined their likelihood surfaces to mimic the analysis in JV14 (with the caveat that all of our mock simulations use the same lensing parameters, appropriate for \mg, rather than eight separate systems).

In the left panel of Figure \ref{fig:mock_constraints}, we combine the likelihood surfaces for eight random mock observations that match the range in chromatic variation displayed in JV14 (in that work, $0.04 < \Delta m_\rmn{max} - \Delta m_\rmn{min} < 0.52$; we use $\Delta m_\rmn{max} - \Delta m_\rmn{min} < 0.6$). These combined observations miss the input $p$ (red star) at greater than the 95 per cent level, instead returning a much lower value in line with the JV14 constraint ($p = 0.8\pm0.2$). In the middle panel of Figure \ref{fig:mock_constraints}, we simply combine eight random observations, imposing no constraints. Although the measured $r_s$ and $p$ are marginally closer to the input values in this case, they still undershoot $p$ at greater than the 95 per cent level. This is due to the prevalence of low chromatic variation observations in our mock sample.

In the right panel of Figure \ref{fig:mock_constraints}, we restrict ourselves to mock simulations meeting the suitability criteria discussed in previous paragraphs ($\Delta m_\rmn{max} - \Delta m_\rmn{min} > 0.4$; $\Delta m_\rmn{red} - \Delta m_\rmn{macro} < 0.3$). Here we accurately recover the input $p$, although there is some evidence in this example combination that we still overestimate the size of the accretion disc $r_s$.

We can make this argument more robust by generating $N$ realisations of eight mock observations each, effectively repeating a JV14-like experiment $N$ times. We can then explore the probability distributions for the recovered $r_s$ and $p$ across the $N$ realisations, using various subsets of the data: completely random, JV14-like (i.e., low chromatic variation), or constrained (i.e., imposing minimum conditions on degree of chromatic variation and convergence to the unmicrolensed baseline).

Given our sample of 200 mock observations, we choose $N=10$. This gives us sufficient mock data points for statistically independent measurements even when we apply restrictions to our sample.

\begin{figure*}
  \includegraphics[width=130mm]{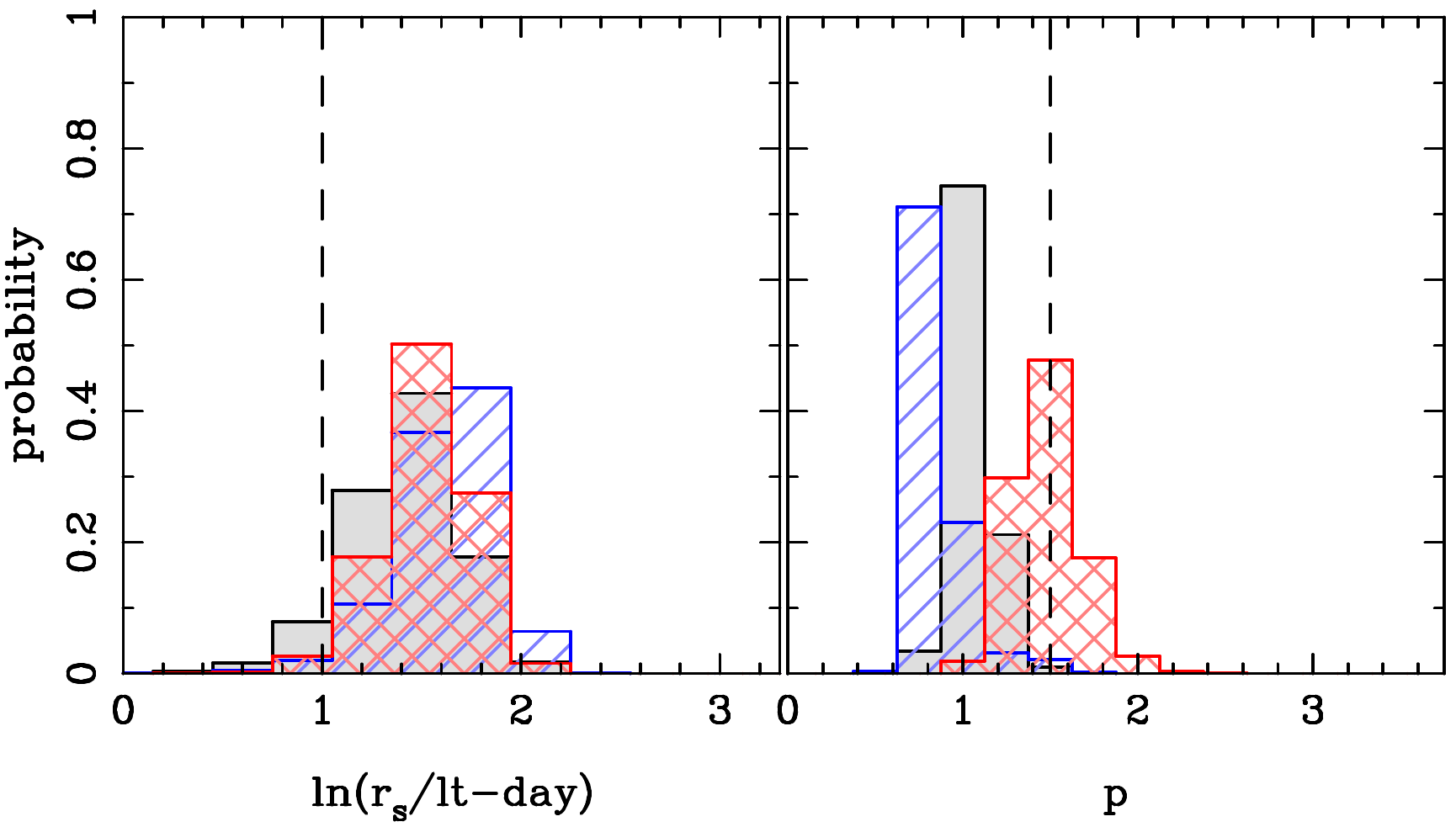}
  \caption{Attempted recovery of input accretion disc parameters (black dashed line) using ten realisations each consisting of eight mock observations drawn from our mock simulation suite. The combined probability distributions are for accretion disc size $r_s$ (left panel) and temperature profile $p$ (right panel).The solid grey histograms represents completely random draws from the full simulation suite. The blue hatched histograms correspond to JV14-like systems, where only low chromatic variation is observed ($\Delta m_\rmn{max} - \Delta m_\rmn{min} < 0.6$). The red histograms correspond to the case where we apply selection criteria to the data: that a minimum chromatic variation of $\Delta m_\rmn{max} - \Delta m_\rmn{min} = 0.4$ is observed, and that the observations roughly converge to the unmicrolensed baseline in the reddest filter ($|\Delta m_\rmn{red} - \Delta m_\rmn{macro}| < 0.3$). Only in this latter case do we correctly recover the input power-law index $p$; all cases tend to overestimate the accretion disc size.}
  \label{fig:histograms}
\end{figure*}

In Figure \ref{fig:histograms} we plot the results of this experiment. The left panel shows the combined probability distribution for accretion disc size $r_s$ of $N=10$ realisations of 8 mock observations each. The right panel shows the distribution for power-law index $p$. The solid grey histograms are drawn randomly from the full 200 mock observations. The blue hatched histograms represent JV14-like datasets, with $\Delta m_\rmn{max} - \Delta m_\rmn{min} < 0.6$. The red cross-hatched histograms are the result of applying minimum criteria to our mock data: that they display chromatic variation $\Delta m_\rmn{max} - \Delta m_\rmn{min} > 0.4$, and roughly converge to the unmicrolensed baseline as $\Delta m_\rmn{red} - \Delta m_\rmn{macro} < 0.3$. 

In all three cases, we systematically overestimate the size of the accretion disc $r_s$, by a factor of $\sim1.5$ to $2.0$. We note that even if selection effects in our data are leading to a factor of 2 over-estimation of accretion disc size, the sizes derived from microlensing observations are typically still larger than expected from thin disc theory.

Constraining the mock observations to match the degree of chromatic variation observed in JV14 leads to a systematic under-estimation of the power-law index $p$. In fact, the result we recover here is identical to the JV14 result: $0.8\pm0.2$. If we randomly select instead from the full suite of mock observations, the recovered value of $p=1.0\pm0.2$ shifts slightly closer to the true value, but still predicts a steeper temperature profile at greater than the $2\sigma$ level. When we apply constraints on both degree of chromatic variation and convergence to the unmicrolensed baseline, we correctly recover the input value of $p=1.5$ (formal measured constraint $p=1.5^{+0.3}_{-0.2}$).

\subsection{Comparison with observed data}
\label{subsec:compare}

There are other observations in the literature which use the same technique as JV14 and this paper to constrain quasar accretion discs. It is interesting to see whether they see similar trends. \citet{rojas+14} presented single-epoch observations (spectra, in this case) of two systems: HE 0047-1756 and SDSS 1155+6346. In both systems, the observed chromatic variation exceeds the rough threshold we find here of $\Delta m_\rmn{max} - \Delta m_\rmn{min} \gtrsim 0.4$. In contrast to JV14, and in line with the general trend observed in this paper, they find $p = 2.3\pm0.8$ for HE 0047-1756 (which displays the most chromatic variation) and $1.5\pm0.6$ for SDSS 1155+6346.

\begin{figure}
  \includegraphics[width=85mm]{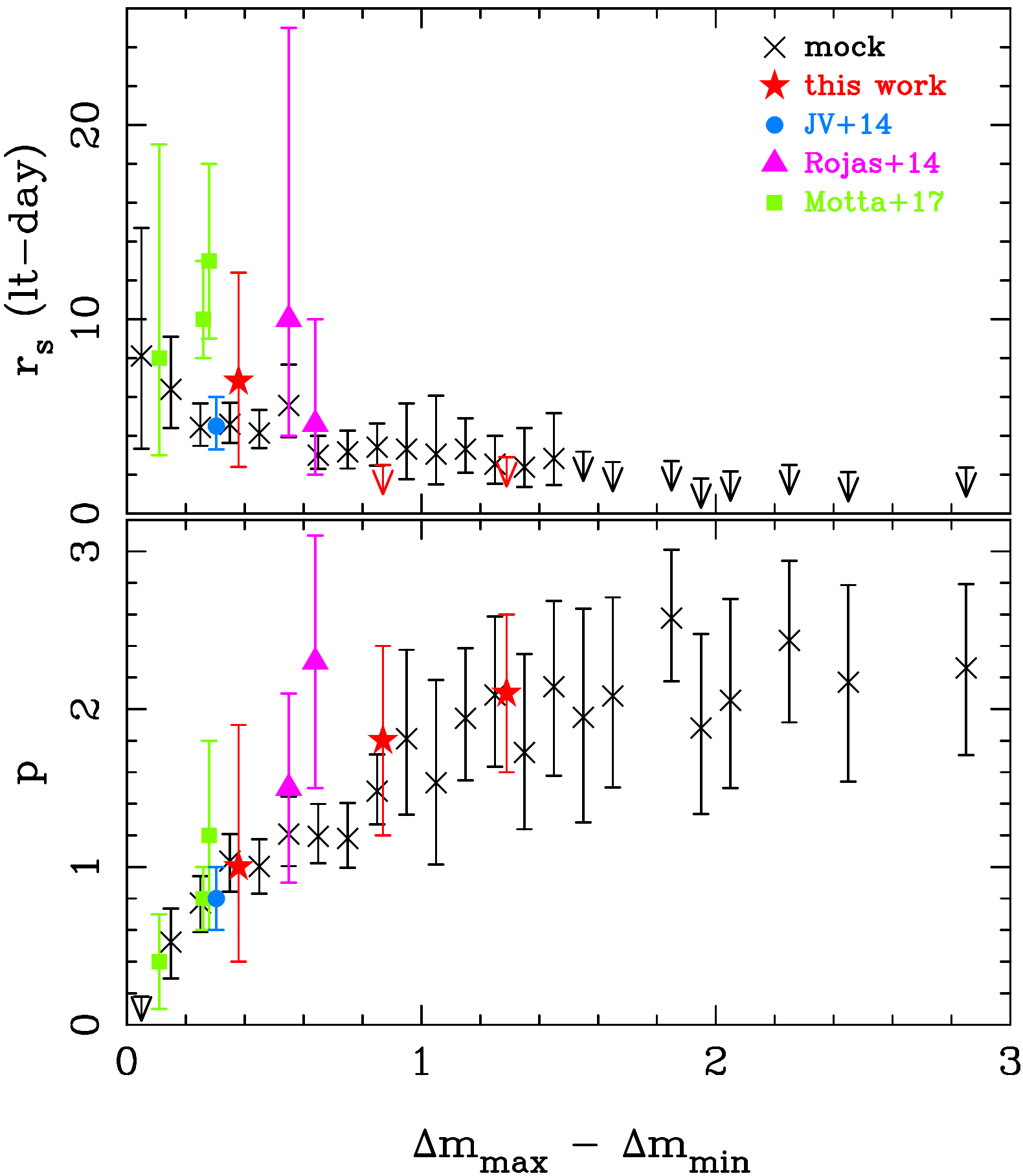}
  \caption{A comparison of mock constraints on $r_s$ (top panel) and $p$ (bottom panel) as a function of chromatic variation, with measurements taken from the literature. Mock data are black crosses, per Figures \ref{fig:mock_rs} and \ref{fig:mock_p}. Red stars: \qsob, \mg, \wfi~(left to right, this work). Light blue circle: a combination of 8 quasars \citep{jv+14}. Magenta triangles: HE 0047-1756 and SDSS 1155+6346 (left to right, \citealt{rojas+14}). Green squares: HE 2149-2745, WFI 2033-4723 (combination of four independent observations), HE 0435-1223 (left to right, \citealt{motta+17}).}
  \label{fig:mock_plus_data}
\end{figure}

\citet{motta+17} analysed spectra for three systems: HE 0435-1223, WFI 2033-4723, and HE 2149-2745. Their observed chromatic variations are low: 0.28, 0.26, and 0.11 respectively (the variation in WFI 2033-4723 is the average of 4 separate epochs of data). As expected, they predict lower values of $p$. In Figure \ref{fig:mock_plus_data} we plot all of these observational constraints: our data, JV14, \citet{rojas+14}, and \citet{motta+17} for $r_s$ and $p$, along with the binned constraints from our mock observations. We note that these are not technically directly comparable -- the observed data are for a range of systems, whereas the mock results are for repeated observations of a single system. Nevertheless, they all follow the same basic trend.

\subsection{Mock observation selection effect discussion}
We propose here that there are two selection effects that lead to systematically underestimating $p$ in single-epoch microlensing simulations: chromatic variation, and convergence to the unmicrolensed baseline. Both of these degeneracies are straight-forward to understand in the context of convolving a power-law accretion disc model with microlensing magnification maps. When chromatic variations are low, the blue (hotter) and red (colder) parts of the accretion disc are similarly affected by microlensing. Preferred solutions then become accretion disc models where size depends only weakly on wavelength ($p$ is low). It is easier to find positions on the magnification map that produce similar microlensing amplitudes if the relative sizes of the blue and red parts of the accretion disc are similar.

An observed flux ratio in the reddest filter that is significantly offset from the unmicrolensed baseline is easier to produce if the reddest emission region is small. Given that we constrain the accretion disc model to be a power-law, if the reddest emission region is small we have less freedom to choose large values for the power-law index $p$ (due to both the constraints on the resolution of our simulations, and physical limits, i.e. the optical/UV accretion disc cannot extend below the innermost stable orbit). 

These findings are related to our earlier observation that small magnitude differences between images provide very little information on the size of the quasar accretion disc \citep{bate+07}. If we insist that the reddest filter roughly converges to the unmicrolensed baseline, then a low degree of chromatic variation means we're learning almost nothing about the accretion disc (size, or temperature profile). If the reddest filter \textit{does not} converge to the unmicrolensed baseline, we may get a tight constraint on size, but $p$ will be artificially driven to low values for the reasons described above.

\subsection{Possible consequences for ensemble measurements}
We have demonstrated the selection effects in the previous section for only one simulated accretion disc, with $r_s = 2.7$ light days and $p=1.5$. The possible consequences of these effects on real ensemble measurements depend on their generalisability. We discuss the three possible cases below: (a) our results are perfectly general; (b) the trends in our results are general; (c) our results are not general.

\subsubsection*{(a) All results are perfectly general} 
In the case where our results are fully general, the selection effects discussed in the previous section hold for all ensemble single-epoch measurements of accretion disc parameters. We can use the following thresholds to ensure that any ensemble of real observed quasars are providing us with accurate measurements of the underlying disc temperature profile:

\begin{itemize}
\item $|\Delta m_\rmn{red} - \Delta m_\rmn{macro}| \lesssim 0.3$, and
\item $\Delta m_\rmn{max} - \Delta m_\rmn{min} \gtrsim 0.4$. 
\end{itemize}

The simplest option for lensed quasar observations that do not meet these requirements is to reject them. If however our results are truly general, then it may be possible to correct for the trends observed in Figures \ref{fig:mock_rs} and \ref{fig:mock_p}. How to do so in an observed dataset, where you have no knowledge of the true accretion disc temperature profile, is not clear.

\subsubsection*{(b) Only trends are general}
In this case, the general trend that low chromatic variation observations lead to underestimating $p$ is true, but the exact shape of the trend depends on some other parameters: perhaps the lens model, or the smooth matter fraction, or the underlying accretion disc temperature profile. In this situation, the thresholds in case (a) might not be universally applicable -- lower chromatic variation observations may be usable in some cases, and misleading in others.

If the shape of these trends depends on external parameters such as the lens model, then we can hopefully account for them by adding a systematic component to our error budgets. If the shape of the trends depends on the underlying accretion disc, the situation is more complicated. Since we have no knowledge of the true temperature profile in a set of real observations, we will not know which of our data are usable. Our best hope in this case is to run suites of mock simulations for various input accretion discs, and determine which ranges of the data return correct results irrespective of the input disc. 

Additionally, in this case it becomes crucially important whether high chromatic variation leads to a corresponding \textit{over-}estimation of $p$. If it does, then any usability thresholds on chromatic variation are likely to be both upper and lower limits; blindly applying a lower limit only might bias us to high $p$, just as using low chromatic variation biases us towards low $p$.

\subsubsection*{(c) Results are not general}
We have demonstrated here that low chromatic variation observations do not lead to correct recovery of the input accretion disc for $p=1.5$. We can imagine, however, that this result is not general -- perhaps the single-epoch method correctly measures accretion disc temperature profiles if they are steep (say, $p\sim1$), but fails to do so if they are shallower (as we have shown for $p=1.5$). This is the worst-case scenario: given a low chromatic variation observation and no knowledge of the true disc temperature profile, we cannot tell whether our measurement is due to the disc itself or a selection effect.

At this stage, we have no reason to suspect that this case is true. We can think of no \textit{a priori} reason why the method would work well for one mock accretion disc, and fail for another. Nevertheless, our current set of mock simulations do not allow us to rule this case out; we include it here for completeness.\\

Based on our mock simulations, \textit{in all three cases described above} we must currently doubt the results of any analyses that make use of ensembles of low chromatic variation observations. If cases (a) or (b) are true, we know that such ensembles are likely to under-estimate $p$ (although, if case (b) is true, we cannot yet be sure where the cutoff in the usefulness of data occurs). If case (c) is true, we (currently) have no way to differentiate between a situation where the true underlying disc has a steep temperature profile and we are correctly measuring it, or it has a shallow temperature profile and we are under-estimating it (as in our mock simulations).

Our analysis for $p=1.5$ accretion discs only does not provide enough information to discern between cases (a), (b), or (c). To do so, we need a suite of mock observations that covers the full range of astrophysically-interesting parameter space: plausible accretion disc temperature profiles, lens models, smooth matter fractions, chromatic variation, and so on. Given the significant computational resources required for such a task, we plan to undertake it in a subsequent paper.

\subsection{Additional simulation caveats}
\label{subsec:caveats}
In our mock simulations we have made two additional simplifying assumptions: we have used only microlensing parameters appropriate for an \mg~like system, and we have assumed that all of the microlensing is in a single image ($A_2$, the saddle point image). We will discuss some implications of these choices below.

\mg~is a high-magnification system ($\mu\sim25$ in the close image pair), with a high density of caustics in its magnification maps. The models used in JV14 have lensed images with a wide range of magnifications, from $\mu\sim0.5$ to $\mu\sim25$. The impact of caustic density on our ability to recover accretion disc parameters with the single-epoch technique has not yet been fully explored. Previous analyses, such as JV14, assume that it makes no difference -- that accretion disc parameters are recovered equally-correctly in low and high magnification systems. We take the same position here. This assumption is circumstantially supported by the similarity between the trend in constraints from our mock observations (using only one set of microlensing parameters) and observations in the literature (see Section \ref{subsec:compare} and especially Figure \ref{fig:mock_plus_data}), however it should be more fully tested in the future.

We have also assumed that all of the microlensing is occurring in the saddle point image $A_2$. It is common in spectroscopic observations of microlensed quasars to find that only one image is strongly affected (see e.g. \citealt{odowd+15}; \citealt{macleod+15}; \citealt{motta+17}), and saddle point images in close image pairs are known to be more strongly effected by microlensing (\citealt{schechter+02}; \citealt{vernardos+14}). Nevertheless, by restricting the microlensing to a single image we do exclude some chromatic signatures from our sample of mock observations. In particular, cases where shorter wavelengths in image $A_1$ are heavily microlensed, leading to an inverted chromatic microlensing curve (this is the case in two of the eight systems presented in JV14).

The mock observations presented here are not intended to be exhaustive. They are, however, all plausible realisations of chromatic microlensing observations in a system like \mg; if the standard single-epoch microlensing technique is free of systematic or selection effects, we should recover the input accretion disc parameters correctly. In future work, we will extend our mock observations to cover all of the plausible microlensing parameter space. This will allow us to robustly demonstrate the impact of these selection effects on single-epoch microlensing measurements in a fully-realistic sample of lensed quasars.\\

\section{Conclusions}
\label{sec:conclusions}

In this paper, we have presented single-epoch HST observations of four gravitationally lensed quasars: \mg; \rxj; \qsob; and \wfi. In each system, we have focussed on close image pairs, where time delays are expected to be negligible. Since we are confident that these images capture the background quasar in the same state, we can use any microlensing-induced chromatic variation to place constraints on the emission regions in the quasar.

Our observations were tuned, as far as was possible, to avoid broad emission lines and therefore provide a clean measurement of microlensing on the quasar accretion discs. Other effects can also masquerade as chromatic microlensing -- most notably differential extinction. We have not attempted to isolate these effects; we assume that the entirety of the chromatic variation is due to microlensing.

Using the observed magnitude differences between close image pairs, we place constraints on the size $r_s$ of the quasar accretion disc at $\lambda_0 =1026$\AA~(rest wavelength) and the power-law index $p$ relating accretion disc radius to rest wavelength. We assume an accretion disc spectral profile of the form $r = r_s (\lambda/\lambda_0)^p$. Our simulation technique broadly follows that of JV14, an extension of \citet{bate+08} and \citet{floyd+09}.

Across our four systems, we find a broad diversity in the measured power-law index $p$, from $p=1.4^{+0.5}_{-0.4}$ in \qsob, to $p = 2.3^{+0.5}_{-0.4}$ in \wfi~at 68 per cent confidence (assuming smooth matter fraction $s=0.8$). This is somewhat at odds with the constraints in JV14, which cluster around $p = 0.8\pm0.2$ (their 68 per cent Bayesian constraint obtained by combining observations from eight systems).

There is a trend in our constraints towards larger values of $p$ when the degree of chromatic variation in our observations is greater. To explore the origin of this trend, we generated a suite of 200 blinded mock observations with a known input accretion disc ($r_s = 2.7$ light days, $p=1.5$). Accretion disc constraints obtained using the standard single-epoch analysis technique on these mock observations display the following trends:

\begin{itemize}
\item In cases where the chromatic variation between the bluest and reddest filters in the mock observations is $\Delta m_\rmn{max} - \Delta m_\rmn{min} \lesssim 0.4$, the single-epoch technique systematically underestimates $p$. Combining multiple mock observations with low chromatic variation in the usual way exacerbates this problem.
\item In cases where the mock flux ratio in the reddest filter differs significantly from the unmicrolensed baseline ($|\Delta m_\rmn{red} - \Delta m_\rmn{macro}|\gtrsim 0.3$), the measured $p$ is once again under-estimated.
\item However, when chromatic variation in the mock observations is sufficiently large, and approximately converges towards the unmicrolensed baseline in the reddest filter, the single-epoch technique correctly recovers input accretion disc power-law index.
\end{itemize}

This leads to the following important conclusions:

\begin{itemize}
\item The combined constraint published in \citet{jv+14} of $p=0.8\pm0.2$, as well as those in \citet{motta+17}, are likely driven by the low degree of chromatic variation in the ensemble of lensed quasars used in those studies.
\item Under the assumption that chromatic variations of $\Delta m_\rmn{max} - \Delta m_\rmn{min} \gtrsim 0.4$ produce clean measurements of the accretion disc temperature profile (implied by our mock observations), we have two systems of interest: \mg~and \wfi. Here, we find very high estimates of $p$: $1.8\pm0.6$ in \mg, $2.3^{+0.5}_{-0.4}$ in \wfi~(assuming $s=0.8$). These correspond to shallower temperature profiles than expected from an SS disc, more in line with \citet{abramowicz+88} slim discs. If these measurements are accurate, they are consistent with the picture that emerges from both microlensing and reverberation mapping analyses, which generically predict larger accretion discs at a given temperature than expected from SS discs.
\end{itemize}

We conclude that there are important selection effects that need to be taken into consideration when using the single-epoch microlensing technique to measure accretion disc temperature profiles. First, this technique should be applied to ensembles of quasars -- measurements from individual objects have large scatter which does not necessarily reflect anything physical in the source. Second, combined results using this technique might be misleading if the chromatic variation is low, or the observations do not roughly converge to the unmicrolensed baseline in the reddest filter. In our mock experiments with $r_s=2.7$ light days and $p=1.5$, we find that the following criteria are sufficient to correctly recover the input values: chromatic variation of $\Delta m_\rmn{max} - \Delta m_\rmn{min} \gtrsim 0.4$, and convergence to the unmicrolensed baseline of $|\Delta m_\rmn{red} - \Delta m_\rmn{macro}|\lesssim 0.3$. Further simulations are required to determine if these thresholds are universally applicable. However, even when these criteria are applied, the technique may still over-estimate accretion disc sizes by roughly a factor of 1.5 to 2.0.

We have conducted only a preliminary exploration of these selection effects here. In particular, we note that the limits quoted in the previous paragraph were estimated using one specific set of microlensing parameters, appropriate for \mg, and assuming that all of the microlensing was occurring in the saddle point image. These effects need to be systematically explored across astrophysically-interesting parameter space if we are to confidently use the single-epoch technique for constraining quasar accretion discs. Detailed analysis of the impact of these effects on mock observations would allow us to construct more appropriate priors on our Bayesian measurements. We will pursue this task in a subsequent paper. 

\section*{Acknowledgements}

We wish to thank the anonymous referee for a thorough and thoughful review, which significantly improved this paper. NFB thanks the STFC for support under Ernest Rutherford Grant ST/M003914/1. Based on observations made with the NASA/ESA Hubble Space Telescope, obtained at the Space Telescope Science Institute, which is operated by the Association of Universities for Research in Astronomy, Inc., under NASA contract NAS 5-26555. These observations are associated with HST Cycle 20 Program ID 12874, PI Floyd. GV is supported through an NWO-VICI grant (project number 639.043.308). RLW acknowledges the support of ARC Grant DP150101727. This work was performed on the gSTAR national facility at Swinburne University of Technology. gSTAR is funded by Swinburne and the Australian Government's Education Investment Fund.

\begin{landscape}
\begin{table}
  \caption{Observed Vega magnitudes of all lensed images and lensing galaxies} \label{tab:fluxes}
  \begin{tabular}{lccccccccc} 
  \hline
  & \multicolumn{9}{c}{Filter} \\
 Object & F410M & F547M & F621M & F689M & F763M & F845M & F105W & F125W & F160W \\
 \hline
 \textbf{\mg} \\ 
 \hspace{12pt} Image $A_1$ & -- & -- & -- & -- & $15.69\pm0.05$ & $14.05\pm0.03$ & -- & $16.60\pm0.01$ & $15.52\pm0.01$ \\
 \hspace{12pt} Image $A_2$ & -- & -- & -- & -- & $16.86\pm0.09$ & $14.98\pm0.05$ & -- & $17.04\pm0.02$ & $15.81\pm0.01$ \\
 \hspace{12pt} Image $B$   & -- & -- & -- & -- & $16.38\pm0.07$ & $14.79\pm0.05$ & -- & $17.55\pm0.02$ & $16.51\pm0.02$ \\
 \hspace{12pt} Image $C$   & -- & -- & -- & -- & $17.24\pm0.11$ & $15.66\pm0.07$ & -- & $18.38\pm0.03$ & $17.31\pm0.02$ \\
 \hspace{12pt} Lens Galaxy & -- & -- & -- & -- & $^a$ & $^a$ & -- & $19.00\pm0.03$ & $18.26\pm0.04$ \\
 \hline
 \textbf{\rxj} \\
 \hspace{12pt} Image $A$ & -- & $16.32\pm0.03$ & $16.00\pm0.04$ & $15.80\pm0.04$ & -- & $15.69\pm0.09$ & -- & $18.70\pm0.04$ & $18.41\pm0.05$ \\
 \hspace{12pt} Image $B$ & -- & $15.75\pm0.02$ & $15.46\pm0.03$ & $15.22\pm0.03$ & -- & $15.08\pm0.05$ & -- & $18.15\pm0.03$ & $17.87\pm0.03$ \\
 \hspace{12pt} Image $C$ & -- & $16.69\pm0.04$ & $16.37\pm0.05$ & $16.14\pm0.05$ & -- & $15.99\pm0.07$ & -- & $18.93\pm0.04$ & $18.63\pm0.05$ \\
 \hspace{12pt} Image $D$ & -- & $16.95\pm0.05$ & $16.60\pm0.06$ & $16.39\pm0.06$ & -- & $16.29\pm0.14$ & -- & $19.28\pm0.04$ & $18.98\pm0.05$ \\
 \hspace{12pt} Lens Galaxy & -- & $^a$ & $^a$ & $^a$ & -- & $^a$ & -- & $18.45\pm0.21$ & $17.77\pm0.32$ \\
 \hline
 \textbf{\qsob} \\
 \hspace{12pt} Image $A$ & -- & -- & $10.86\pm0.01$ & -- & $10.14\pm0.01$ & $9.58\pm0.01$ & $15.47\pm0.01$ & $14.95\pm0.02$ & $14.50\pm0.01$ \\
 \hspace{12pt} Image $B$ & -- & -- & $11.35\pm0.01$ & -- & $10.59\pm0.01$ & $9.98\pm0.01$ & $15.74\pm0.01$ & $15.12\pm0.02$ & $14.61\pm0.01$ \\
 \hspace{12pt} Image $C$ & -- & -- & $11.43\pm0.01$ & -- & $10.69\pm0.01$ & $10.13\pm0.01$ & $16.01\pm0.01$ & $15.53\pm0.02$ & $15.07\pm0.01$ \\
 \hspace{12pt} Image $D$ & -- & -- & $14.56\pm0.03$ & -- & $13.79\pm0.03$ & $13.22\pm0.04$ & $19.00\pm0.04$ & $18.45\pm0.11$ & $17.88\pm0.05$ \\
 \hspace{12pt} Lens Galaxy & -- & -- & $^a$ & -- & $^a$ & $^a$ & $19.23\pm0.05$ & $18.71\pm0.09$ & $18.11\pm0.08$ \\
 \hline
 \textbf{\wfi} \\
 \hspace{12pt} Image $A_1$ & $15.21\pm0.03$ & $15.04\pm0.02$ & -- & $14.38\pm0.02$ & $14.06\pm0.02$ & $13.83\pm0.03$ & -- & $16.32\pm0.01$ & $15.76\pm0.01$ \\
 \hspace{12pt} Image $A_2$ & $14.29\pm0.02$ & $14.78\pm0.02$ & -- & $14.37\pm0.02$ & $14.13\pm0.02$ & $13.95\pm0.03$ & -- & $16.60\pm0.01$ & $16.14\pm0.02$ \\
 \hspace{12pt} Image $B$   & $15.57\pm0.04$ & $16.01\pm0.03$ & -- & $15.56\pm0.04$ & $15.28\pm0.03$ & $15.09\pm0.06$ & -- & $17.74\pm0.04$ & $17.20\pm0.04$ \\
 \hspace{12pt} Image $C$   & $15.72\pm0.05$ & $16.17\pm0.04$ & -- & $15.75\pm0.05$ & $15.49\pm0.04$ & $15.32\pm0.07$ & -- & $17.92\pm0.04$ & $17.46\pm0.04$ \\
 \hspace{12pt} Lens Galaxy & $^a$ & $^a$ & -- & $^a$ & $^a$ & $^a$ & -- & $19.92\pm0.29$ & $18.62\pm0.11$ \\
 \hline
\multicolumn{10}{l}{$^a$ Not detected in our data}\\
\end{tabular}
\end{table}

\begin{table}
  \caption{Observed magnitude differences used in the current microlensing analysis} \label{tab:ratios}
  \begin{tabular}{lcccccccccc} 
  \hline
  & & \multicolumn{9}{c}{Observed wavelength (\AA)} \\
 System & Images & 4109 & 5447 & 6219 & 6876 & 7612 & 8436 & 10552 & 12486 & 15369 \\
 \hline
 \mg & $A_2 - A_1$ & -- & -- & -- & -- & $1.17\pm0.11$ & $0.93\pm0.06$ & -- & $0.44\pm0.02$ & $0.29\pm0.01$ \\ 
 \rxj & $B - A$ & -- & $-0.57\pm0.04$ & $-0.54\pm0.05$ & $-0.58\pm0.05$ & -- & $-0.61\pm0.10$ & -- & $-0.55\pm0.05$ & $-0.54\pm0.06$ \\
 \qsob & $B - A$ & -- & -- & $0.49\pm0.01$ & -- & $0.45\pm0.01$ & $0.40\pm0.01$ & $0.27\pm0.01$ & $0.17\pm0.02$ & $0.11\pm0.01$ \\
 \wfi & $A_2 - A_1$ & $-0.92\pm0.03$ & $-0.26\pm0.02$ & -- & $-0.01\pm0.03$ & $0.07\pm0.02$ & $0.12\pm0.04$ & -- & $0.28\pm0.02$ & $0.38\pm0.02$ \\
 \hline
\end{tabular}
\end{table}
\end{landscape}

\begin{landscape}

\begin{table}
  \caption{Observed microlensing amplitudes, defined as $\Delta m_\rmn{micro} = \Delta m_\rmn{obs} - \Delta m_\rmn{macro}$} \label{tab:dm_micro}
  \begin{tabular}{lccccccccccc} 
  \hline
  & & \multicolumn{9}{c}{Observed wavelength (\AA)} \\
 System & Images & $\Delta m_\rmn{macro}$ & 4109 & 5447 & 6219 & 6876 & 7612 & 8436 & 10552 & 12486 & 15369 \\
 \hline
 \mg & $A_2 - A_1$ & $0.09\pm005^a$ & -- & -- & -- & -- & $1.08\pm0.12$ & $0.84\pm0.08$ & -- & $0.35\pm0.05$ & $0.20\pm0.05$ \\ 
 \rxj & $B - A$ & $-0.74\pm0.10^b$ & -- & $0.17\pm0.11$ & $0.20\pm0.11$ & $0.16\pm0.11$ & -- & $0.13\pm0.14$ & -- & $0.19\pm0.11$ & $0.20\pm0.12$ \\
 \qsob & $B - A$ & $-0.08\pm0.12^c$ & -- & -- & $0.57\pm0.12$ & -- & $0.53\pm0.12$ & $0.48\pm0.12$ & $0.35\pm0.12$ & $0.25\pm0.12$ & $0.19\pm0.12$ \\
 \wfi & $A_2 - A_1$ & $0.14\pm0.25^d$ & $-1.06\pm0.25$ & $-0.40\pm0.25$ & -- & $-0.15\pm0.25$ & $-0.08\pm0.25$ & $-0.03\pm0.25$ & -- & $0.14\pm0.25$ & $0.24\pm0.25$ \\
 \hline
 \multicolumn{10}{l}{$^a$ \citet{minezaki+09}; \citet{macleod+13}}\\
 \multicolumn{10}{l}{$^b$ \citet{jackson+15}}\\
 \multicolumn{10}{l}{$^c$ \citet{patnaik+99}}\\
 \multicolumn{10}{l}{$^d$ Obtained from macro-modelling, rather than from observed IR or radio data. Errors in $\Delta m_\rmn{macro}$ added to observed errors in quadrature (see Sections \ref{subsubsec:isolate} and \ref{subsubsec:sims}).}\\
\end{tabular}
\end{table}
\end{landscape}



\bibliographystyle{mnras}
\bibliography{nbate_hstmicrolensing} 

\bsp	
\label{lastpage}
\end{document}